\documentclass{article}

\usepackage{PRIMEarxiv}

\usepackage[utf8]{inputenc} % allow utf-8 input
\usepackage[T1]{fontenc}    % use 8-bit T1 fonts
\usepackage{hyperref}       % hyperlinks
\usepackage{url}            % simple URL typesetting
\usepackage{tikz}
\usepackage{xcolor}
\usetikzlibrary{positioning, calc, decorations.pathreplacing}
\usepackage{booktabs}       % professional-quality tables
\usepackage{amsfonts}       % blackboard math symbols
\usepackage{lscape}
\usepackage{fontawesome5}
\usepackage{colortbl}
\usepackage{array, multirow} % Add this in the preamble
\usepackage{float}
\usepackage{svg}
\usepackage{xstring}
\usepackage{nicefrac}       % compact symbols for 1/2, etc.
\usepackage{microtype}      % microtypography
\usepackage{lipsum}
\usepackage{fancyhdr}       % header
\usepackage{graphicx}       % graphics
\graphicspath{{media/}}     % organize your images and other figures under media/ folder

\title{Surveying Uncertainty Representation: A Unified Model for Cyber-Physical Systems
}

\author{
  Johannes Mäkelburg \\
  School of Computation, Information and Technology \\
  Technical University of Munich \\
  Heilbronn, Germany\\
  \texttt{johannes.maekelburg@tum.de} \\
   \And
  Diego Perez-Palacin\\
  Linnaeus University \\
  Växjö, Sweden \\
  and \\
  Karlsruhe Institute of Technology \\
  Karlsruhe, Germany \\
  \texttt{diego.perez-palacin@lnu.se} \\
  \And
  Raffaela Mirandola\\
  Karlsruhe Institute of Technology \\
  Karlsruhe, Germany \\
  \texttt{raffaela.mirandola@kit.edu} \\
  \And
  Maribel Acosta \\
  School of Computation, Information and Technology \\
  Technical University of Munich \\
  Heilbronn, Germany\\
  \texttt{maribel.acosta@tum.de} \\
}

\begin{document}
\maketitle

\begin{abstract}
Cyber-Physical Systems (CPS) operate in dynamic environments, leading to different types of uncertainty.
 This work provides a comprehensive review of uncertainty representations and categorizes them based on the dimensions used to represent uncertainty.
 Through this categorization, key gaps and limitations in existing approaches are identified. 
 To address these issues, a Conceptual Model of Uncertainty Representations in CPS is introduced, integrating and extending existing models.
 Its applicability is demonstrated through examples from the automotive domain, showing its effectiveness in capturing and structuring uncertainty in real-world scenarios.
\end{abstract}

% keywords can be removed

\section{Introduction}~\label{sec:introduction}
% Topic Introduction:
% Introduction CPS 
Cyber-physical systems (CPS) integrate computational components with physical processes through embedded computers that control the system via feedback loops~\cite{lee2008cyber, lee2015past}.
This integration enables continuous interaction and mutual influence between the physical and computational components~\cite{lee2008cyber, lee2015past, baheti2011cyber}. 
With the rapid advancement of digitalization, CPS have become a fundamental component of modern technological infrastructures.
Today, nearly all technological infrastructures can be considered CPS, with prominent examples including automotive systems~\cite{chakraborty2016automotive}, smart grids~\cite{yu2016smart}, and healthcare~\cite{haque2014review}. 
% Why CPS are important for real-world application 
A key advantage of CPS lies in their ability to simulate physical processes within a cyber environment, which accelerates technological development, improves system efficiency, and reduces overall costs~\cite{baheti2011cyber}.
CPS achieve these benefits by enabling real-time access to information, supporting predictive maintenance, facilitating predefined decision-making, and optimizing operational processes~\cite{oztemel2020literature}. 
However, CPS also introduce significant challenges.
They are highly vulnerable to security threats~\cite{oztemel2020literature, reddy2015security,} and different types of uncertainties~\cite{lee2015past}.  The literature offers diverse interpretations of uncertainty, from a complete lack of understanding to insufficient data and the gap between the information needed to complete a task and the current information \cite{acosta2022uncertainty,perez2014uncertainties,ramirez2012taxonomy,weyns2023towards}.
A general definition, that can be adopted also for CPS is provided by Walker~\cite{walker2003defining}, where uncertainty is seen as \emph{any deviation from the unachievable ideal of
completely deterministic knowledge of the relevant system}. 
%These deviations may result in diminished confidence in the findings, stemming from the perception that they could be partial, ambiguous, imprecise, untrustworthy, indecisive, or potentially erroneous.
%\jm{Include definition of Uncertainty here}
%Uncertainty in Cyber-Physical Systems refers to insufficient knowledge about the system’s current state
%In the domain of CPS, Uncertainty is defined as the lack of knowledge about the current conditions of the system. %state of the system
% Why Uncertainty is crucial for CPS 
This issue is exacerbated in CPS, which, unlike traditional software systems, must operate in dynamic and often unpredictable environments where uncertainties arise from multiple sources, including sensor inaccuracies, environmental variations, and unforeseen interactions between system components \cite{mahdavi2017classification}. 
Managing these uncertainties is particularly crucial for CPS due to the tight coupling between computational and physical processes.
Since CPS rely on the precise coordination of these components, uncertainties can propagate throughout the system, affecting functionality and overall reliability.
In safety-critical domains %such as autonomous driving and healthcare, 
such disruptions can have severe consequences, including system failures and safety risks.
%The complexity of CPS is further amplified by its interdisciplinary nature, as it integrates domain-specific concepts from each included domain. 
%Differences between these domains include an additional source of uncertainty, especially in the interaction between components from different domains.  
Moreover, CPS operate across multiple domains, where their interdisciplinary nature introduces additional sources of uncertainty, particularly between components of different domains.
These uncertainties, along with the dynamic and unpredictable nature of CPS, underscore the need for a systematic approach to managing uncertainty throughout the entire system lifecycle.
Effectively addressing them requires integrating uncertainty considerations from the earliest stages of software design.
Addressing these uncertainties effectively requires incorporating the notion of uncertainty from the beginning of software design.
A design approach that inherently accounts for uncertainties allows CPS development to be guided by principles and methodologies that explicitly consider uncertainties. 
%Keywords from the Journal Scope:
%Cyber-Physical Systems (CPS) \cite{lee2008cyber}
%Literature Review of the State-of-the-Art Uncertainty Representations in CPS
%Framework/ Taxonomy 
%Interdisciplinary (Uncertainty in interdisciplinary Domain)
%Future Directions 
%\paragraph{Problem Statement}
% Challenges for Representing Uncertainty in CPS 
% Lack of Unified Terminology and Missing Dimensions due to the new emerging Technology in CPS 
%\paragraph{Problem Statement}
%While CPS provide significant advantages in automation, efficiency, and real-time decision-making, their operation is inherently subject to various uncertainties.%, including sensor inaccuracies, environmental changes, and unforeseen system interactions. 
%Their interdisciplinary nature further amplifies these challenges, as uncertainties arise not only from technical limitations but also from human decision-making and regulatory constraints. 
%Existing approaches to uncertainty representation are fragmented, leading to inconsistencies across domains and a growing divergence in terminology.
%A structured framework is needed to systematically represent uncertainty, ensuring the reliability, safety, and adaptability of CPS in real-world applications.
\paragraph{Related Uncertainty Surveys}
%R: Here, we need to cite related surveys 
%Uncertainty representation has been widely explored across different domains, each with its own focus and scope.
%While numerous surveys offer valuable insights into specific aspects of uncertainty, they often provide a limited perspective, either by focusing on a single domain or addressing only a subset of uncertainty dimensions. 
%In contrast, our work aims to present a more comprehensive view, particularly in the context of Cyber-Physical Systems (CPS). 
%This section reviews key surveys in the field, highlighting their contributions and clarifying how our approach differs.
Uncertainty representation has been extensively studied across various domains, with numerous surveys providing valuable insights. 
However, these works often focus on specific aspects of uncertainty, such as a particular model type, a single uncertainty dimension, or a restricted application domain.
In contrast, our work aims to present a more comprehensive perspective, particularly tailored to the challenges of uncertainty in CPS. 

We categorize related surveys into three groups: (1) those that focus on uncertainty representation in specific model types, (2) surveys that address particular categories of uncertainty representation, and (3) studies that explore uncertainty within CPS.
This section reviews key contributions in each category, highlighting their scope, limitations, and distinctions from our approach.
%\jm{Include Short Title for each Related Survey Category}

\textit{(1) Model-Specific Uncertainty Surveys:}
Among existing surveys, many focus on uncertainty representation within specific model types~\cite{guo2022survey, gawlikowski2023survey, wang2024uncertainty, fakour2024structured, chen2024uncertainty, shorinwa2024survey}.
%Several surveys focus exclusively on uncertainty representations within a specific type of model.
Guo et al.~\cite{guo2022survey} present a comprehensive survey on uncertainty reasoning and quantification, bridging classical belief-based frameworks with deep learning to enhance AI decision-making. 
However, while their work focuses on integrating classical uncertainty models into deep learning, our study provides a broader exploration of uncertainty representations in CPS.
Similarly, Gawlikowski et al.~\cite{gawlikowski2023survey} concentrate exclusively on uncertainty estimation in neural networks.
Wang et al.~\cite{wang2024uncertainty} and Chen et al.~\cite{chen2024uncertainty} both explore uncertainty representations in Graph Neural Networks (GNN). 
While Wang et al.~\cite{wang2024uncertainty} focus solely on uncertainty in GNNs, Chen et al.~\cite{chen2024uncertainty} take a broader approach by also considering uncertainty quantification in Probabilistic Graph Models.  
%Gawlikowski et al.~\cite{gawlikowski2023survey} provide an overview of uncertainty estimation in neural networks, restricting their focus to this particular model type.
%provide an overview of uncertainty estimation in neural networks (only consideration of one potential model type (neural networks))
%Similarly, Wang et al.~\cite{wang2024uncertainty} limit their survey to only exploring uncertainty representations in Graph Neural Networks. 
%\cite{wang2024uncertainty} - Uncertainty Representation in Graph Neural Networks (Different scope, only one type of possible model (GNNs))
Fakour et al.\cite{fakour2024structured} conduct an extensive survey on uncertainty representation by reviewing uncertainty categories, uncertainty sources, and methods to quantify uncertainty, limiting their scope to Machine Learning Processes. 
%But again, they also limit their review to Machine Learning Processes.
%\cite{fakour2024structured} - review of the literature in the various facets that uncertainty is enveloped in the ML process, defining uncertainty categories, uncertainty sources, uncertainty quantification theory (Only consideration of one possible model (ML-models))
%Chen et al.~\cite{chen2024uncertainty} survey uncertainty quantification in machine learning but focus exclusively on graphical models, including Graph Neural Networks and Probabilistic Graphical Models, covering model architectures, training, and inference.
Shorinwa et al.~\cite{shorinwa2024survey} extensively review existing methods for uncertainty quantification in large language models. 
%no interaction between domains 
%\cite{li2012dealing} -  survey about uncertainty processing activities in diverse fields (look at uncertainty representations in different fields but not at uncertainty representations in mixed fields like CPS)

\textit{(2) Category-Specific Uncertainty Surveys:}
Other surveys focus on a particular category of uncertainty representation, limiting their scope to a single category rather than a comprehensive representation~\cite{zio2013literature, quan2019survey, keith2021survey, acar2021modeling, haugen2023representation, cuzzolin2024uncertainty}.
Zio et al.~\cite{zio2013literature} focus on reviewing methods to represent uncertainty analysis and output uncertainty, particularly in the context of modeling the effects of uncertainty when models are used for decision-making. 
Quan et al.~\cite{quan2019survey} focus in their survey on a specific domain of CPS - smart power grids with high renewable energy penetration – and survey how uncertainty is quantified and managed for wind power generation. 
Keith et al.~\cite{keith2021survey} extend this focus by not only considering representations of the effect of uncertainty in decision-making but also in optimization.
Haugen et al.~\cite{haugen2023representation}  examine various representations of uncertainty effects in the domain of power market models for operational planning and forecasting.  
Cuzzolin~\cite{cuzzolin2024uncertainty} surveys different theoretical frameworks for uncertainty measurement, restricting its focus to a single category of uncertainty representation.
Acar et al.~\cite{acar2021modeling} review uncertainty effect modeling practices in design optimization of structural and multidisciplinary systems under uncertainties.

%effect 
%\cite{zio2013literature} - review of methods to represent uncertainty analysis, output uncertainty when models are used for decision-making (scope is engineering applications and focus on output uncertainty)
%\cite{keith2021survey} - approaches for representing uncertainty in both decision-making and optimization (modeling the effect of uncertainty is the focus)
%\cite{haugen2023representation} - representation of uncertainty in power market models for operational planning and forecasting (represents only one category of our understanding of uncertainty representation - modeling the effect of uncertainty)
%\cite{cuzzolin2024uncertainty} - Survey on Uncertainty Measurements (represents only one category of our understanding of uncertainty representation)
%\cite{acar2021modeling} - review of the uncertainty treatment practices in design optimization of structural and multidisciplinary systems under uncertainties ()
Kamal et al.~\cite{kamal2021recent} interpret uncertainty representation in their survey on uncertainty visualization as a representation of a quantification and visualization approach.  
In contrast, our survey takes a broader view, considering uncertainty representation beyond quantification and visualization.
%\cite{kamal2021recent} - Survey on Uncertainty Visualization (interpretation of Uncertainty representation as Quantification- and Visualization Approach) 

\textit{(3) Method-Specific Uncertainty Surveys:}
A few surveys address uncertainty in CPS, but they often focus on specific methods rather than offering a structured framework for uncertainty representation.
Li et al.~\cite{li2012dealing} review multidisciplinary uncertainty processing activities across various fields, analyzing different types of uncertainties and their handling approaches. 
However, their survey focuses on how uncertainty is managed within individual disciplines without considering interactions between them.
Tao et al.~\cite{tao2020uncertainty} conducted a survey on uncertainty management in CPS, examining state-of-the-art qualitative, quantitative, and visualization methods for managing uncertainty.
While their survey explores multiple dimensions of uncertainty, our work takes a broader perspective by encompassing a wider range of uncertainty dimensions in CPS.

\paragraph{Contributions}
In this work, we provide a comprehensive literature review of uncertainty representations in CPS.
We systematically categorize existing representations based on their categories used to represent uncertainties. %the categories used to represent uncertainties. 
Through this categorization, we conduct an in-depth analysis that identifies key gaps and limitations in the current literature.
%Based on this, we conducted an analysis, which led to the identification of some gaps and limitations in the current literature.
To fill these gaps and overcome these limitations, we propose a Conceptual Model of Uncertainty Representations in CPS, which harmonizes existing models and extends them by incorporating missing categories necessary for a more comprehensive representation of uncertainty. 
To evaluate the applicability of the proposed model, we apply it to multiple examples from the automotive domain, demonstrating its effectiveness in capturing and structuring uncertainty in real-world CPS scenarios.

Our contributions can be summarized as follows:
\begin{enumerate}
    \item A structured literature review that categorizes and analyzes existing uncertainty representations in CPS, identifying key gaps and limitations.  
    \item A harmonized and extended Conceptual Model of Uncertainty Representations that integrates and refines existing models while introducing missing categories.  
    \item A demonstration of the model’s applicability through case studies in the automotive domain, showing its effectiveness in capturing and structuring uncertainty.  
\end{enumerate}
This work advances the understanding of uncertainty representation in CPS and provides a foundation for future research on more expressive and structured modeling approaches.

\paragraph{Paper Structure} 
The remainder of this paper is structured as follows. 
Section~\ref{sec:literatureReview} reviews existing work on uncertainty representation in CPS, categorizing different modeling approaches and identifying key contributions and gaps. 
%This analysis forms the foundation for the development of our conceptual model.
Section~\ref{sec:proposedFramework} introduces our harmonized and extended Conceptual Model, which addresses the gaps identified in the literature. 
%It refines existing uncertainty representations to better capture the characteristics of uncertainty in CPS.
Section~\ref{sec:example} demonstrates the applicability of our conceptual model using an Autonomous Vehicle as a case study. 
%We illustrate how different uncertainty categories manifest in this system and how our model structures them.
Section~\ref{sec:conclusion} concludes with a summary and future research directions
%Section~\ref{sec:conclusion} concludes the paper by summarizing our key findings and discussing potential directions for future research. 

%\input{chapters/Background}

%\input{tables/literatureOverview}
\section{Overview of the related Literature}
\label{sec:literatureReview}

%\paragraph{Literature Search}
%R: 1) We need to make explicit the way in which we retrieved the papers we cite.
The objective of this literature review is to identify and analyze the most relevant uncertainty representations for the domain of CPS. 
Rather than conducting an exhaustive survey, we focus on key contributions that provide a representative overview of the current state of the art.

Our selection prioritized peer-reviewed journal articles and conference proceedings, particularly those that address uncertainty representation beyond a single modeling paradigm or application domain. 
%Priority was given to peer-reviewed journal articles and conference proceedings, with a focus on works that discuss uncertainty representation beyond a single modeling paradigm or application domain. 
Works that were limited to highly specialized techniques without broader relevance to CPS were excluded.  
%Papers that were narrowly scoped to specific techniques without broader relevance to CPS were excluded.
This selection process allowed us to construct a representative collection of literature that underpins our analysis of uncertainty representation in CPS.
To identify relevant work, we searched major academic databases, including \textit{IEEE Xplore}, \textit{ACM Digital Library}, \textit{SpringerLink}, and \textit{Google Scholar}. 
Our search queries combined keywords such as \textit{Cyber-Physical Systems}, \textit{Uncertainty Representation}, \textit{Uncertainty Modeling}, and \textit{Uncertainty Quantification}. 
Additional relevant works were identified through backward and forward citation tracking to ensure the inclusion of foundational and influential contributions.
%We identified additional relevant works through backward and forward citation tracking, ensuring that foundational and influential contributions were included. 

%The objective of the literature review is to identify and analyze the most relevant uncertainty representations for the domain of Cyber-Physical Systems.
%not to achieve an exhaustive survey of all existing models. 
%Instead, we focus on identifying and analyzing the most prominent and relevant uncertainty representations that can be applied to the context of Cyber-Physical Systems. 
%Rather than offering an exhaustive survey, we focus on key contributions that provide a representative overview of the current state of the art.
%By concentrating on these key representations, we aim to provide a representative and insightful overview of the current state of the art without diving into a comprehensive review of all the existing literature.

This selection analyses papers from various relevant domains, focusing on the domains of Software Engineering, Cyber-Physical Systems, and Coupled Models.
These domains were selected based on their contributions to understanding and managing uncertainty in CPS.
A complete list of all the literature included in the review can be found in Table \ref{tab:literatureOverview}.
The subsequent sections of this chapter categorize the literature according to the dimensions or categories used to classify and represent uncertainty in the literature.

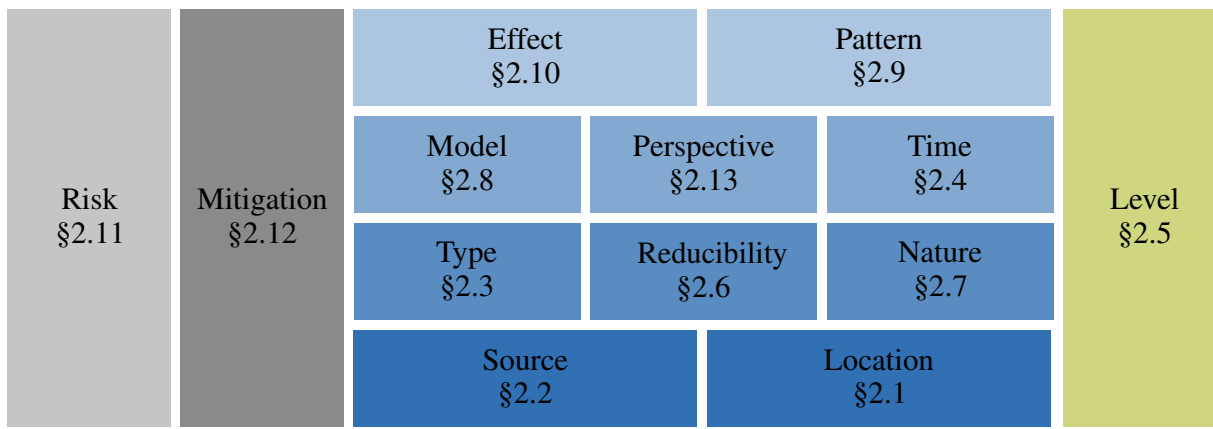
\begin{figure}[t!]
    \centering
    \resizebox{\textwidth}{!}{
    \begin{tikzpicture}[
    smallbox/.style={minimum width=1.4cm, minimum height=1cm, align=center, draw=none, inner sep=5pt},
    largebox/.style={minimum width=1.4cm, minimum height=4.745cm, align=center, draw=none, inner sep=5pt},
    mediumbox/.style={minimum width=3.9cm, minimum height=1.1cm, align=center, draw=none, inner sep=5pt},
    box/.style={minimum width=2.57cm, minimum height=1.1cm, align=center, draw=none, inner sep=5pt},
    node distance=0.1cm % Adjust the node distance here
    ]
    % Define color
    \definecolor{tumwebblue}{HTML}{3070b3}
    \definecolor{tumlightblue}{HTML}{64A0C8}
    \definecolor{tumgrey}{HTML}{8a8a8a}
    \definecolor{tumgreen}{HTML}{A2AD00}
    \definecolor{tumorange}{HTML}{E37222}
    \definecolor{tumblue}{HTML}{005293}%
    \definecolor{tumdark}{HTML}{000000}
    \definecolor{tumpurple}{HTML}{7E57C2}
    \definecolor{tumcoral}{HTML}{E55B54}
    
    \definecolor{color1}{RGB}{229, 204, 255}   % Light Purple
    \definecolor{color2}{RGB}{189, 224, 255}   % Light Blue
    \definecolor{color3}{RGB}{255, 223, 186}   % Light Peach
    \definecolor{color4}{RGB}{255, 239, 186}   % Light Yellow
    \definecolor{color5}{RGB}{255, 204, 204}   % Light Pink
    \definecolor{color6}{RGB}{204, 255, 204}   % Light Green
    \definecolor{color7}{RGB}{204, 255, 255}   % Light Teal
    % First column
    \node[largebox, fill=tumgrey!50] (risk) {\begin{minipage}{1.5cm}\centering Risk \\ \S \ref{sec:risk}\end{minipage}};
    % Second column
    \node[largebox, fill=tumgrey, right=of risk] (mitigation) {\begin{minipage}{1.5cm}\centering Mitigation \\ \S \ref{sec:mitigation} \end{minipage}};
    %\node[largebox, fill=tumpurple!40] (risk) {\begin{minipage}{1.5cm}\centering Risk \\ \S \ref{sec:risk}\end{minipage}};
    % Second column
    %\node[largebox, fill=tumpurple!60, right=of risk] (mitigation) {\begin{minipage}{1.5cm}\centering Mitigation \\ \S \ref{sec:mitigation} \end{minipage}};
    % Third column
    \node[box, fill=tumwebblue!60, right=of mitigation, yshift=0.605cm] (model) {\begin{minipage}{2cm}\centering Model \\ \S \ref{sec:modelType} \end{minipage}};
    \node[mediumbox, fill=tumwebblue!40, above=of model, xshift= 0.66cm] (effect) {\begin{minipage}{2cm}\centering Effect \\ \S \ref{sec:effect} \end{minipage}};
    \node[box, fill=tumwebblue!60, right=of model] (perspective) {\begin{minipage}{2cm}\centering Perspective \\ \S\ref{sec:perspective} \end{minipage}};
    \node[mediumbox, fill=tumwebblue!40, right=of effect] (pattern) {\begin{minipage}{2cm}\centering Pattern \\ \S \ref{sec:pattern} \end{minipage}};
    % Thrid Row
    \node[box, fill=tumwebblue!80, below=of model] (type) {\begin{minipage}{1.5cm}\centering Type \\ \S \ref{sec:uncertaintyTypes} \end{minipage}};
    \node[box, fill=tumwebblue!80, right=of type] (reducibility) {\begin{minipage}{1.5cm}\centering Reducibility \\ \S \ref{sec:reducibility} \end{minipage}};
    \node[box, fill=tumwebblue!80, right=of reducibility] (nature) {\begin{minipage}{1.4cm}\centering Nature \\  \S \ref{sec:nature} \end{minipage}};
    \node[box, fill=tumwebblue!60, right=of perspective] (time) {\begin{minipage}{2cm}\centering Time \\ \S \ref{sec:time} \end{minipage}};
    % Fourth Row
    \node[mediumbox, fill=tumwebblue, below=of type, xshift=0.66cm] (source) {\begin{minipage}{1.5cm}\centering Source \\ \S \ref{sec:source} \end{minipage}};
    \node[mediumbox, fill=tumwebblue, right=of source] (location) {\begin{minipage}{1.5cm}\centering Location \\ \S \ref{sec:location} \end{minipage}};
    % Right column
    \node[largebox, fill=tumgreen!50, right=of time, yshift=-0.61cm] (level) {\begin{minipage}{1.5cm}\centering Level \\ \S \ref{sec:level} \end{minipage}};
    %\node[largebox, fill=tumpurple!80, right=of time, yshift=-0.61cm] (level) {\begin{minipage}{1.5cm}\centering Level \\ \S \ref{sec:level} \end{minipage}};
    \end{tikzpicture}
    }
    \caption{Overview of different categories representing Uncertainty across various frameworks.}
    \label{fig:uncertaintyCategories}
\end{figure}

\subsection{Location}~\label{sec:location}

According to Walker et al.~\cite{walker2003defining}, the location of uncertainty refers to the point within a complex model where uncertainty manifests. This helps identify where the uncertainty that affects the outcome is created in the model. Walker et al. outline several typical locations of uncertainty that can apply to most models: \textit{Context}, \textit{Model}, \textit{Inputs}, \textit{Parameters}, and \textit{Outcomes}. \textit{Context} refers to the system's boundaries and how fully it represents the real world. \textit{Model} uncertainty includes both conceptual model uncertainty, which relates to the variables and relationships, and technical uncertainty, which concerns the computer implementation. \textit{Inputs} pertain to the description of the reference system and external forces driving its changes. \textit{Parameters} refer to uncertainty in the data and methods used to calibrate the model parameters. Lastly, \textit{Outcome} uncertainty is the difference between model outcomes and true values that is significant to decision-makers.
Chipman et al.~\cite{chipman2015coverage} build on Walker et al.'s framework, categorizing uncertainty locations as \textit{Inputs}, \textit{Parameters}, and \textit{Models}. \textit{Input} uncertainties typically stem from a lack of knowledge about initial conditions or unforeseen scenarios. \textit{Parameter} uncertainties arise when accurate values are unknown, though statistical knowledge or value ranges may be available. \textit{Model} uncertainty refers to uncertainties in the representation of the existing system.

%\textcolor{red}{The three next sentences have been slightly modified}

The taxonomy by Perez-Palacin et al.~\cite{perez2014uncertainties} builds on Walker et al.~\cite{walker2003defining}, tailoring it to the uncertainties in the models used by self-adaptive systems. It identifies three types of uncertainty location: \textit{Context}, \textit{Model structural}, and \textit{Input}. The \textit{Context uncertainty} concerns the decided model boundaries with respect to the real world where the system operates, while \textit{Model structural uncertainty} represents how accurately the model characterizes the modeled subset of the real world. \textit{Input}, also called parameter, refers to uncertainty concerning the actual values of input variables or calibrating methods.
Pelz et al.~\cite{pelz2021mastering} introduce \textit{System Design} as a category to identify uncertainty locations, including \textit{Model} uncertainty, \textit{Structural} uncertainty similar to Perez-Palacin et al.~\cite{perez2014uncertainties}, and \textit{Data} uncertainty arising from incomplete, unclear, or insufficient data.
Acosta et al.~\cite{acosta2022uncertainty} combine the understandings of Pelz et al.~\cite{pelz2021mastering} and Perez-Palacin et al.~\cite{perez2014uncertainties} into a new notion called \textit{Locus}, which specifies which model elements are affected and their corresponding location in the real world. The locus categories are \textit{Parameters}, \textit{Models}, \textit{Analysis}, and \textit{Decision Making}. \textit{Parameter} includes uncertainties related to all types of parameters that characterize a model, while uncertainty related to the model definition, formalism selection, boundaries, and structure are combined in the category \textit{Models}. The category \textit{Analysis} comprises uncertainty from evaluating design decisions through model-based methods, and uncertainties in the category \textit{Decision-Making} relate to the decision-making process when different options are available or when the decision is based on incomplete information and only estimates of the values of interest are available.

Mahdavi-Hezavehi et al.~\cite{mahdavi2017classification} categorize the locations of uncertainty within a system. Potential locations include \textit{Environment}, \textit{Model}, \textit{Adaptation functions}, \textit{Goals}, \textit{Managed systems}, and \textit{Resources}. \textit{Environment} encompasses the execution context and human interactions affecting the system.
%\diego{(Something weird happened in the previous sentence. There is a \texttt{\textbackslash textit} that should be only ``text'', but maybe some other text has dissapeared.)}
\textit{Model} refers to the various conceptual models representing the system. \textit{Adaptation functions} correspond to the functionalities performed as part of MAPE-K~\cite{kephart2003vision}. \textit{Goals} group the specification, modeling, and modification of system goals. \textit{Managed Systems} organize all application-specific monitoring and adaptation systems. \textit{Resources} include the essential factors and components required for the self-adaptive system to operate normally.

Other works do not further categorize uncertainty locations but provide constructs to represent them. 
Zhang et al.~\cite{zhang2016understanding} propose a Belief Model where the location of uncertainty is defined in the Uncertainty Model as the specific place where uncertainty occurs. 
Bandyszak et al.~\cite{bandyszak2020orthogonal} present an Uncertainty Ontology and use the term \textit{Observation Point} to document the artifacts containing uncertainty, which serve as locations where the system can detect uncertainty. Each observation point may be associated with multiple data sources, representing multiple uncertainty locations.
\subsection{Source}
\label{sec:source}

In our selected literature, Ramirez et al.~\cite{ramirez2012taxonomy} provide the first description of Uncertainty Sources, identifying 26 distinct uncertainty sources within dynamically adaptive systems. 
These sources are organized across three critical phases of the model lifecycle—\textit{Requirement Level}, \textit{Design-Time}, and \textit{Run-Time}—providing a lifecycle-oriented framework. 

Building on this foundation, the FORMS reference architecture~\cite{weyns2010forms} has served as a lens to explore uncertainty in self-adaptive systems. 
FORMS breaks a self-adaptive system (SAS) into four different components: the User, the Environment, the Base-Level -- which includes the main functionalities of the system -- and the Meta-Level, which manages the Base-Level's behavior using the MAPE-K feedback control loop~\cite{kephart2003vision}. 
Based on interactions between these components, Esfahani et al.~\cite{esfahani2013uncertainty} identify uncertainty sources such as \textit{simplified assumptions}, \textit{model drift}, \textit{noise}, \textit{parameters in future operation}, \textit{human in the loop}, \textit{objectives}, \textit{decentralization}, \textit{context}, and \textit{cyber-physical systems}. 
%Some of these sources are later reused by Perez-Palacin et al.~\cite{perez2014uncertainties} and Mahdavi-Hezavehi et al.~\cite{mahdavi2017classification}.
Rather than proposing new sources, Perez-Palacin et al.~\cite{perez2014uncertainties} classify these sources into a taxonomy designed to highlight similarities. 
This approach enables the application of shared mitigation methods for similar types of uncertainty.
The work by Mahdavi-Hezavehi et al.~\cite{mahdavi2017classification} defines uncertainty sources as circumstances that cause a system to deviate from its expected behavior, similar to \textit{model drift} from Esfahani et al.~\cite{esfahani2013uncertainty}.

Unlike previous works, Bastin et al.~\cite{bastin2013managing} categorize the uncertainty into two primary origins: \textit{Model Input} and \textit{Model Structure}, each subdivided into more specific sources. 
\textit{Model Input} uncertainties include \textit{Measurement}, \textit{Representativity}, \textit{Sensor Model}, and \textit{Transmission Uncertainties}. 
\textit{Model Structure} uncertainties encompass \textit{Mechanism}, \textit{Representation}, \textit{Parameter}, and \textit{Numerical Uncertainties}.

The concept of \textit{Indeterminacy}, introduced by Zhang et al.\cite{zhang2016understanding} in their Belief Model, adds another layer to the understanding of uncertainty sources. 
Here, uncertainty is framed as a lack of confidence in belief statements due to insufficient knowledge, with contributing factors termed \textit{Indeterminacy Sources}. 
Chatterjee et al.\cite{chatterjee2020toward} adopts this framework. 

Expanding on these ideas, Bandyszak et al.~\cite{bandyszak2020orthogonal} incorporate the notion of uncertainty sources into their Uncertainty Ontology, which they refer to as \textit{Uncertainty Rationale}. 
Their subsequent work~\cite{bandyszak2022uncertainty} introduces the ECDC model to address uncertainty in real-time computing. % focusing on core aspects like \textit{Execution Platforms}, \textit{Communication Infrastructure}, \textit{Data Processing}, and \textit{Coordination}.
This model distinguishes various kinds of uncertainty based on four core concepts: \textit{Execution platform}, which includes the hardware and operating system for running real-time embedded software; \textit{Communication infrastructure}, referring to resources for data exchange and coordination between devices; \textit{Data Processing}, encompassing uncertainties in processing data from sensors or other technical devices; and \textit{Coordination}, addressing uncertainties from collaboration between autonomous systems.
This model emphasizes the operational contexts in which uncertainty arises. 

Further broadening the scope, Acosta et al.~\cite{acosta2022uncertainty} distinguish between system-related and environment-related uncertainty sources, offering an abstract perspective that complements the more detailed frameworks. 
Lastly, Asmat et al.~\cite{asmat2023uncertainty}, in a literature review specific to CPS, synthesize earlier approaches, including Zhang et al.’s Belief Model and Bandyszak et al.’s Uncertainty Ontology. Their taxonomy organizes uncertainty causes into three categories: \textit{Human Behavior}, \textit{Natural Processes}, and \textit{Technological Processes}, underscoring the diverse origins of uncertainty in CPS.

\subsection{Uncertainty Types}
\label{sec:uncertaintyTypes}

The concept of \textit{Uncertainty Types} is included in the model of Zhang et al.~\cite{zhang2016understanding} encompasses five categories: Occurrence, Environment, Content, Geographical Location, and Time Uncertainty. 
This model has influenced subsequent work, such as Chatterjee et al.~\cite{chatterjee2020toward} and the PSUM-Metamodel~\cite{PSUM2023}, which adopt the concept of uncertainty types. 

Troya et al.~\cite{troya2021uncertainty} further explore uncertainty in software models, reviewing various methods for representing it, including Zhang et al.'s Uncertainty Model. 
Their analysis expands the classification into a taxonomy comprising six uncertainty types: \textit{Spatiotemporal}, \textit{Measurement}, \textit{Occurrence}, \textit{Design}, \textit{Behavior}, and \textit{Belief Uncertainty}. 
This taxonomy reflects the growing need to address uncertainty in increasingly complex systems and provides new perspectives on its categorization.

One uncertainty type that both Zhang et al.~\cite{zhang2016understanding} and Troya et al.~\cite{troya2021uncertainty} have in common is \textit{Occurrence Uncertainty}.
Zhang et al. describe this as the lack of confidence in the occurrence of events mentioned in a Belief Statement, whereas Troya et al. interpret it as uncertainty about the existence of an entity. 
Zhang et al. also define Geographical Location Uncertainty and Time Uncertainty, representing a lack of confidence in spatial and temporal aspects, respectively. 
Troya et al., however, combine these dimensions into a single type, Spatiotemporal Uncertainty, highlighting a more integrated view of uncertainty related to space and time.

Beyond these shared aspects, Zhang et al.~\cite{zhang2016understanding} identify distinct types of uncertainty, for example, environment and content uncertainty. 
\textit{Environment Uncertainty} refers to a Belief Agent's lack of confidence in the surroundings of a physical system.  
\textit{Content Uncertainty} pertains to uncertainty in the details of a Belief Statement's content. 

Complementary, Troya et al.~\cite{troya2021uncertainty} introduce additional types of uncertainty that address specific aspects of system modeling. 
\textit{Measurement Uncertainty} relates to the possible states or outcomes of a measurement, typically associated with probabilities. 
\textit{Design Uncertainty} captures ambiguity in system design decisions, including uncertainties about user requirements, operating conditions, and potential solutions. 
\textit{Behavior Uncertainty} involves the unpredictability of system or environmental behavior, encompassing actions, motivations, timing, and parameters. 
Finally, \textit{Belief Uncertainty} represents uncertainty in any statement made about the system or its environment, reflecting a broad, overarching category that applies to various aspects of modeling and analysis.

\subsection{Time}
\label{sec:time}

Ramirez et al.~\cite{ramirez2012taxonomy} propose a template for representing uncertainty in systems, including a \textit{Classification} category that links uncertainty to the phase of the system where it occurs:  
the \textit{Requirement Level}, which captures uncertainties during the setting of requirements; 
\textit{Design-Time}, where uncertainties emerge during the system's development process; 
and \textit{Run-Time}, which encompasses uncertainties encountered after deployment. 
Acosta et al.~\cite{acosta2022uncertainty} and Mahdavi-Hezavehi et al.~\cite{mahdavi2017classification} also include a temporal categorization, distinguishing only between \textit{Design-Time} and \textit{Run-Time}.
These categorizations capture how uncertainties can manifest and evolve in the lifecycle of a system.

Zhang et al.~\cite{zhang2016understanding} and Chatterjee et al.~\cite{chatterjee2020toward} introduce an additional perspective by examining the duration of uncertainty and describing its ``lifetime'':  
\textit{Temporal Uncertainty} occurs within a specific time interval, and \textit{Persistent Uncertainty},  remains unresolved until specific actions or events take place. 
This classification provides a dynamic view of uncertainty, capturing its temporal behavior. % and highlighting how it interacts with system processes over time.

%Ramirez et al.~\cite{ramirez2012taxonomy} created a template for describing different types and sources of uncertainty. 
%One category called \textit{Classification} assigns each uncertainty to the phase in which it occurs. 
%The phases include \textit{Requirement Level}, which refers to the time when development requirements are established, \textit{Design-Time}, describing the phase during development, and \textit{Run-Time}, the time once the system is deployed.
%Both Acosta et al.~\cite{acosta2022uncertainty} and Mahdavi-Hezavehi et al.~\cite{mahdavi2017classification} differentiate only between \textit{Design-Time} and \textit{Run-Time} regarding the existence of uncertainty. 

%Zhang et al.~\cite{zhang2016understanding} and Chatterjee et al.~\cite{chatterjee2020toward} describe in their Uncertainty Model the lifetime of uncertainty as the time interval during which it occurs.
%Either uncertainty occurs temporally for a specific time interval, \textit{Temporal Uncertainty} or is persistent until it is resolved, \textit{Persistent Uncertainty}.
\subsection{Level}
\label{sec:level} 

Walker et al.~\cite{walker2003defining} propose a spectrum of knowledge levels to categorize uncertainties, ranging from deterministic knowledge to total ignorance in five levels: 
\textit{determinism}, 
\textit{statistical uncertainty}, 
\textit{scenario uncertainty}, 
\textit{recognized ignorance}, and 
\textit{total ignorance}. 
In this context, \textit{statistical uncertainty} refers to uncertainty that can be characterized using statistical terms, while \textit{scenario uncertainty} arises when only a range of possible outcomes is identifiable, with limited understanding of the mechanisms leading to them.
Building on Walker’s framework, Esfahani et al.~\cite{esfahani2013uncertainty} introduce the term \textit{Spectrum of Uncertainty} and redefine the endpoints as \textit{certainty} and \textit{ignorance}, drawing from Aughenbaugh~\cite{aughenbaugh2006managing}. 
They add two intermediate states: \textit{Current Information}, representing the present level of knowledge, and \textit{Complete Information}, marking the theoretical limit of knowable facts. 
%The spectrum is further divided: the interval from Certainty to Complete Information is labeled \textit{Irreducible Uncertainty}, while the segment between Complete and Current Information is termed \textit{Reduced Uncertainty}.
Mahdavi-Hezavehi et al.~\cite{mahdavi2017classification} align with Walker’s and Esfahani’s definitions but focus primarily on distinguishing \textit{statistical uncertainty} and \textit{scenario uncertainty}, adopting Walker’s original terminology.

%\jm{Mention here already the link of scenario and statistical uncertainty to Uncertainty Effect?}
%Esfahani et al.~\cite{esfahani2013uncertainty} adopt the definition from Walker et al.~\cite{walker2003defining} but use the term \textit{Spectrum of Uncertainty}. 
%They use and extend upon terms introduced by Aughenbaugh~\cite{aughenbaugh2006managing}, picking \textit{certainty} and \textit{ignorance} as endpoints instead of \textit{determinism} and \textit{total ignorance}. 
%To differentiate levels of uncertainty, they introduce the terms \textit{Current Information}, which represents the current state of knowledge, and \textit{Complete Information}, marking the point where all knowable information is attained, which marks the point at which everything knowable is known. 
%The interval between Certainty and Current Information is called Imprecision, with Complete Information placed within this interval. 
%Therefore, this interval can be divided into parts: the section from Certainty to Complete Information is labeled Irreducible Uncertainty, while the range from Complete Information to Current Information is termed Reduced Uncertainty.

%Mahdavi-Hezavehi et al.\cite{mahdavi2017classification} refers to the shared definition of the Level/Spectrum of Uncertainty from Walker et al.~\cite{walker2003defining} and Esfahani et al.~\cite{esfahani2013uncertainty} but only differentiates between \textit{scenario uncertainty} and \textit{scenario uncertainty} while using the definition provided by Walker et al.~\cite{walker2003defining}. 

Furthermore, Zhang et al.~\cite{zhang2016understanding} differentiate three \textit{Levels of Occurrence} of uncertainty in their Belief Model: the Application Level, resulting from events within the CPS application; the Infrastructure Level, resulting from interactions among physical units; and the Integration Level, which results from interactions between uncertainties either within a level or across different levels.

Perez-Palacin et al.~\cite{perez2014uncertainties} adopt Armour’s \textit{orders of ignorance} framework~\cite{armour2000five}, which offers a broader perspective on uncertainty. 
This approach categorizes uncertainty into five levels: the 0th order (complete certainty), the 1st order (known uncertainty, such as statistical uncertainty), the 2nd order (gaps in knowledge), the 3rd order (unawareness of these gaps and a lack of methods to address them), and the 4th order (meta uncertainty), which reflects uncertainty about the existence of the previous levels.
Walker et al.~\cite{walker2003defining} consolidate the higher orders under the term \textit{Total Ignorance}.

%Perez-Palacin et al.~\cite{perez2014uncertainties} on the other end follow the more general approach from Armour~\cite{armour2000five} the \textit{orders of ignorance}. 
%This approach contains the following five levels: 
%The 0th order represents a complete lack of uncertainty, essentially full knowledge.
%The 1st order represents a known lack of knowledge or 'known uncertainty.' % \jm{including statistical uncertainty, scenario uncertainty, and acknowledged ignorance, as defined by Walker et al.~\cite{walker2003defining}.}
%At the 2nd order, there is a lack of knowledge and  awareness about this gap. 
%The 3rd order adds another layer, where there is a lack of awareness and no process available to recognize this unawareness. 
%Finally, the 4th order, known as Meta Uncertainty, reflects uncertainty regarding the nature or existence of these various orders of uncertainty.
%%\jm{Walker et al.~\cite{walker2003defining} combine these last three levels under the term \textit{Total Ignorance}.}\\

\begin{landscape}
\begin{table}[p]
\centering
\caption{Overview of used Literature and its Uncertainty Categories}
\label{tab:literatureOverview}
\rowcolors{2}{white}{gray!20}
\begin{tabular}{|p{5cm}|>{\centering\arraybackslash}m{0.8cm}|>{\centering\arraybackslash}m{1.1cm}|>{\centering\arraybackslash}m{0.8cm}|>{\centering\arraybackslash}m{0.8cm}|>{\centering\arraybackslash}m{0.9cm}|>{\centering\arraybackslash}m{0.7cm}|>{\centering\arraybackslash}m{0.7cm}|>{\centering\arraybackslash}m{0.6cm}|>{\centering\arraybackslash}m{0.7cm}|>{\centering\arraybackslash}m{0.6cm}|>{\centering\arraybackslash}m{1.3cm}|>{\centering\arraybackslash}m{1.5cm}|}
\hline
\rowcolor{white}
\footnotesize{\textbf{Paper}} & \footnotesize{\textbf{Source}} & \footnotesize{\textbf{Location}} & \footnotesize{\textbf{Nature}} & \footnotesize{\textbf{Model}} & \footnotesize{\textbf{Pattern}} & \footnotesize{\textbf{Effect}} & \footnotesize{\textbf{Level}} & \footnotesize{\textbf{Time}} & \footnotesize{\textbf{Types}}& \footnotesize{\textbf{Risk}}& \footnotesize{\textbf{Mitigation}}& \footnotesize{\textbf{Perspective}}\\ \hline
\scriptsize{A classification framework of uncertainty in architecture-based self-adaptive systems with multiple quality requirements~\cite{mahdavi2017classification}} & \multirow{3}{*}{\centering$\checkmark$} & \multirow{3}{*}{\centering$\checkmark$}  & \multirow{3}{*}{\centering$\checkmark$} &  &  &  &  \multirow{3}{*}{\centering$\checkmark$} & \multirow{3}{*}{\centering$\checkmark$} & & & & \\ \hline
\scriptsize{Orthogonal uncertainty modeling in the engineering of cyber-physical system~\cite{bandyszak2020orthogonal}} & \multirow{2}{*}{\centering$\checkmark$} & \multirow{2}{*}{\centering$\checkmark$} &  &  & \multirow{2}{*}{\centering$\checkmark$} &  \multirow{2}{*}{\centering$\checkmark$} &  &  &  &  &\multirow{2}{*}{\centering$\checkmark$}& \\ \hline
\scriptsize{Mastering uncertainty in mechanical engineering~\cite{pelz2021mastering}} &  & \multirow{2}{*}{\centering$\checkmark$} &  &  &  & \multirow{2}{*}{\centering$\checkmark$} &  &  & &  & & \\ \hline
\scriptsize{Defining uncertainty: a conceptual basis for uncertainty
management in model-based decision support~\cite{walker2003defining}} &  & \multirow{3}{*}{\centering$\checkmark$} & \multirow{3}{*}{\centering$\checkmark$} &  &  &  & \multirow{3}{*}{\centering$\checkmark$} &  & &  & & \\ \hline
\scriptsize{Uncertainties in the modeling of self-adaptive systems: A taxonomy and an example of availability evaluation~\cite{perez2014uncertainties}} & \multirow{3}{*}{\centering$\checkmark$} & \multirow{3}{*}{\centering$\checkmark$} & \multirow{3}{*}{\centering$\checkmark$} &  &  &  & \multirow{3}{*}{\centering$\checkmark$} &  & &  & & \\ \hline
\scriptsize{Uncertainty Theories for Real-Time System~\cite{bandyszak2022uncertainty}} & $\checkmark$ &  &  &  &  &  &  &  & &  & & \\ \hline
\scriptsize{Uncertainty in self-adaptive software systems~\cite{esfahani2013uncertainty}} & \checkmark &  &  &  &  &  & \checkmark &  & &  & & \\ \hline
\scriptsize{A taxonomy of uncertainty for dynamically adaptive sys-
tems~\cite{ramirez2012taxonomy}} & \multirow{2}{*}{\centering$\checkmark$} &  &  &  &  & \multirow{2}{*}{\centering$\checkmark$} &  & \multirow{2}{*}{\centering$\checkmark$} & &  & \multirow{2}{*}{\centering$\checkmark$}& \\ \hline
\scriptsize{Managing uncertainty in integrated environmental modelling: The UncertWeb framework~\cite{bastin2013managing}} & \multirow{2}{*}{\centering$\checkmark$} &  &  &  &  &  &  &  & &  & &  \\ \hline
\scriptsize{Uncertainty in coupled models of cyber-physical systems~\cite{acosta2022uncertainty}} & \multirow{2}{*}{\centering$\checkmark$} & \multirow{2}{*}{\centering$\checkmark$} &  & \multirow{2}{*}{\centering$\checkmark$} &  & \multirow{2}{*}{\centering$\checkmark$} &  & \multirow{2}{*}{\centering$\checkmark$} &  &   & &  \\ \hline
\scriptsize{Uncertainty representation in software models: a survey~\cite{troya2021uncertainty}} &  &  &  &  &  &  &  &  &  \multirow{2}{*}{\centering$\checkmark$}  &  & &  \\ \hline
\scriptsize{Coverage of uncertainties in cyber-physical systems~\cite{chipman2015coverage}} &  & \multirow{2}{*}{\centering$\checkmark$} &  &  & \multirow{2}{*}{\centering$\checkmark$} & \multirow{2}{*}{\centering$\checkmark$} &  &  & &  & & \\ \hline
\scriptsize{“Incorporating measurement uncertainty into OCL/UML primitive datatypes~\cite{bertoa2020incorporating}} &  &  &  & \multirow{2}{*}{\centering$\checkmark$} &  &  &  &  & &  & & \\ \hline
\scriptsize{Understanding uncertainty in cyber-physical systems: a conceptual model~\cite{zhang2016understanding}} & \multirow{2}{*}{\centering$\checkmark$} & \multirow{2}{*}{\centering$\checkmark$} &  &  & \multirow{2}{*}{\centering$\checkmark$} & \multirow{2}{*}{\centering$\checkmark$} & \multirow{2}{*}{\centering$\checkmark$} & \multirow{2}{*}{\centering$\checkmark$} & \multirow{2}{*}{\centering$\checkmark$}& \multirow{2}{*}{\centering$\checkmark$}  & & \\ \hline
\scriptsize{Toward modeling and verification of uncertainty in cyber-physical systems~\cite{chatterjee2020toward}} & \multirow{2}{*}{\centering$\checkmark$} & \multirow{2}{*}{\centering$\checkmark$} &  &  & \multirow{2}{*}{\centering$\checkmark$} & \multirow{2}{*}{\centering$\checkmark$} &  & \multirow{2}{*}{\centering$\checkmark$} & \multirow{2}{*}{\centering$\checkmark$}& \multirow{2}{*}{\centering$\checkmark$}  & & \\ \hline
\scriptsize{Uncertainty handling in cyber-physical systems: State-of-the-art approaches, tools, causes, and future directions~\cite{asmat2023uncertainty}} & \multirow{2}{*}{\centering$\checkmark$} &  &  &  &  &  &  &  &  &  & \multirow{2}{*}{\centering$\checkmark$} & \\ \hline
\scriptsize{Precise Semantics for Uncertainty Modeling (PSUM), Version 1.0~\cite{PSUM2023}} &  &  & \multirow{2}{*}{\centering$\checkmark$} &  & \multirow{2}{*}{\centering$\checkmark$} & \multirow{2}{*}{\centering$\checkmark$} & \multirow{2}{*}{\centering$\checkmark$} &  & \multirow{2}{*}{\centering$\checkmark$} & \multirow{2}{*}{\centering$\checkmark$} & & \multirow{2}{*}{\centering$\checkmark$}\\ \hline
\end{tabular}
\end{table}
\end{landscape}
\subsection{Reducibility}
\label{sec:reducibility}

The PSUM-Metamodel~\cite{PSUM2023} introduces the concept of Reducibility Level to classify uncertainty based on its potential for reduction.
This dimension provides a way to represent how much uncertainty can be mitigated through additional information or analysis.
Fully Reducible Uncertainty refers to situations where uncertainty, although present, can be entirely reduced until full certainty is achieved. 
Partially Reducible Uncertainty describes uncertainty that can only be reduced to a limited extent. 
Finally, Irreducible Uncertainty denotes cases where no additional certainty can be attained, and the uncertainty cannot be reduced.

\subsection{Nature}
\label{sec:nature}

Walker et al.~\cite{walker2003defining}, Perez-Palacin et al.~\cite{perez2014uncertainties}, Mahdavi-Hezavehi et al.~\cite{mahdavi2017classification} and the PSUM-Metamodel~\cite{PSUM2023} all use the nature of uncertainty to describe uncertainty. 
The works by Pelz et al.~\cite{pelz2021mastering} and Troya et al.~\cite{troya2021uncertainty} also explore this dimension, though not as an explicit category.

A consistent theme across these works is the distinction between epistemic and aleatory uncertainty. 
\textit{Epistemic Uncertainty} emerges due to the imperfection of knowledge or understanding of a phenomenon or system. 
\textit{Aleatory Uncertainty}, results from the inherent variability of a phenomenon or system. 
As a result, Mahdavi-Hezavehi et al.~\cite{mahdavi2017classification} refer to it as \textit{Variability Uncertainty} instead. 

\subsection{Type of Model}
\label{sec:modelType}

Uncertainty representation can be tackled in models with different properties, e.g., software or mathematical models. 
Bertoa et al.~\cite{bertoa2020incorporating} present methods for incorporating measurement uncertainty into OCL/UML primitive data types, such as Boolean, Integer, and String. 
Perez-Palacin et al.~\cite{perez2014uncertainties} use UML diagrams and Markovian models to represent uncertainty in software behavior and performance. 
These works establish the foundations for uncertainty-aware software engineering models. 
Pelz et al.\cite{pelz2021mastering} use a mechanical engineering perspective by employing mathematical models, such as functions and ordinary differential equations, to address uncertainty in physical processes. 
In the domain of CPS, Acosta et al.~\cite{acosta2022uncertainty} comprises models from both mechanical~\cite{pelz2021mastering} and software engineering~\cite{perez2014uncertainties}. 
This work emphasizes the importance of integrating techniques that account for uncertainty across diverse components of these systems.

\subsection{Pattern}
\label{sec:pattern}

The characterization by Chipman et al.~\cite{chipman2015coverage} distinguishes two fundamental patterns of uncertainty: static and dynamic. 
Static uncertainties describe unknown variables that are constant over time, whereas dynamic uncertainties must be determined for every use since they change to a given law. 
Consequently, a dynamic uncertainty can be defined as a sequence of multiple static uncertainties.

Building on this distinction, Zhang et al.~\cite{zhang2016understanding} propose a framework called the \textit{Occurrence Pattern of Uncertainty}, which categorizes how uncertainty arises. 
Their model identifies two primary types: random uncertainty, which lacks any discernible pattern, and temporal uncertainty, which is characterized by its adherence to temporal patterns. 
Temporal patterns are further refined into two categories: systematic patterns, which are mathematically predictable and can be either persistent (ongoing indefinitely) or periodic (recurring at regular intervals), and aperiodic patterns, which occur irregularly and are classified as sporadic (occasional) or transient (temporary). 
The PSUM-Metamodel~\cite{PSUM2023} adopts and integrates these pattern types from Zhang et al.'s framework.
Chatterjee et al.~\cite{chatterjee2020toward} expand on Zhang et al.'s model by introducing a third primary type of uncertainty: spatial. 
Spatial uncertainties depend on location rather than time and differ from temporal uncertainties in that they only manifest systematically, either persistently or periodically.

Further refinement of these concepts is found in Bandyszak et al.'s Uncertainty Ontology~\cite{bandyszak2020orthogonal}, which refers to the Occurrence Pattern of Uncertainty as the \textit{Activation Condition}. 
This term describes the specific circumstances under which uncertainties are triggered during runtime.

\subsection{Uncertainty Effect}
\label{sec:effect}

Ramirez et al.~\cite{ramirez2012taxonomy} introduce the concept of \textit{Impact} in their template to describe uncertainties. 
This dimension specifically addresses how uncertainty influences the design or execution of a system. 
Chipman et al.~\cite{chipman2015coverage} further differentiate between probabilistic and non-deterministic models to capture the nature of uncertainty, and then distinguish between continuous and discrete models to address its structural representation. 

Zhang et al.~\cite{zhang2016understanding} present a  \textit{Measure Model} with three categories to describe the effects of uncertainty: 
\textit{Ambiguity}, involves measuring uncertainty in terms of ambiguous or imprecise observations;  
\textit{Probability}, applies probabilistic measures to quantify uncertainty;  
\textit{Vagueness} uses fuzzy methods or qualitative measures to represent uncertainty. 
Both Chatterjee et al.~\cite{chatterjee2020toward} and Bandyszak et al.~\cite{bandyszak2020orthogonal} adopt and adjust this model, with Bandyszak et al. framing it under the term \textit{Uncertainty Effect}.% in their Uncertainty Ontology.

Pelz et al.\cite{pelz2021mastering} offer another perspective on representing the effects of uncertainty by introducing a tiered framework. 
They distinguish between cases where the effect of uncertainty is known versus unknown. 
When unknown, it is labeled as \textit{ignorance}. When known, further differentiation is made: if the probability of the effect is known, it is called \textit{stochastic uncertainty}; if not, it is called \textit{incertitude}. 
Acosta et al.~\cite{acosta2022uncertainty} align with this classification. However, they diverge in their third classification, suggesting an uncharacterized form of uncertainty to represent cases where further delineation is unnecessary or impractical.

Finally, the PSUM-Metamodel~\cite{PSUM2023} frames the effect of uncertainty as a direct consequence of uncertainty within a belief statement. 
This perspective emphasizes how uncertainty can impact the interpretation and reliability of such statements, such as when misinterpretation arises due to ambiguous or incomplete information.

%\begin{comment}

%Ramirez et al.~\cite{ramirez2012taxonomy} introduce a category called \textit{Impact} in their template for describing uncertainties, which explains the effects that uncertainty may have on the design or execution of the system.

%Chipman et al.~\cite{chipman2015coverage} apply two distinct characterizations to capture the behavior of uncertainty. 
%First, they differentiate between probabilistic and non-deterministic models. 
Then, they distinguish between continuous and discrete models.

\subsection{Uncertainty Risk}
\label{sec:risk} 

Zhang et al.~\cite{zhang2016understanding} highlight  that not all uncertainties pose the same level of risk. 
To address this, they employ ISO 31000~\cite{iso31000} to categorize uncertainties into four risk levels: Low, Medium, High, and Extreme. 
This classification considers both the probability of occurrence and the potential impact, with the latter being assessed through a Risk Matrix~\cite{lansdowne1999risk}.
Building on this foundation, Chatterjee et al.~\cite{chatterjee2020toward} and the PSUM-Metamodel~\cite{PSUM2023} integrate this classification system into their frameworks.

\subsection{Uncertainty Mitigation}
\label{sec:mitigation}

Ramirez et al.~\cite{ramirez2012taxonomy} include a category called \textit{Mitigation Strategies} in their template for uncertainties in dynamically adaptive systems. 
This category enumerates available techniques to resolve specific sources of uncertainty.
Building on this work, the Uncertainty Ontology~\cite{bandyszak2020orthogonal} features a similar concept called \textit{Uncertainty Mitigation}.
It comprises methods aimed at proactively preventing uncertainties from arising, rather than merely addressing them once they occur. 

Asmat et al.~\cite{asmat2023uncertainty} contribute further by proposing a taxonomy based on their literature review of uncertainty sources. 
Their taxonomy identifies key dimensions where tools and approaches can be applied to manage uncertainty. These dimensions include Physical Units -- subdivided into Hardware and Software Units -- Heterogeneous Networks, External Entities such as human behavior, and the Physical Environment. 
These dimensions provide a structure for addressing uncertainties across diverse system components.

\subsection{Uncertainty Perspective}
\label{sec:perspective}

The PSUM Metamodel~\cite{PSUM2023} introduces the concept of the perspective, which categorizes uncertainty as \textit{subjective} or \textit{objective} based on the viewpoint of observing agents.  
A subjective perspective arises when uncertainty is shaped by the observations and reasoning processes of the agent, reflecting an interpretation that varies across different observers. In contrast, an objective uncertainty is independent of any specific observer's influence or interpretation.
\subsection{Identified Gaps and Limitations}
\label{sec:gapsLimitations}
\begin{figure*}[t!]
\centering
    \resizebox{0.95\textwidth}{!}{%
    \includegraphics{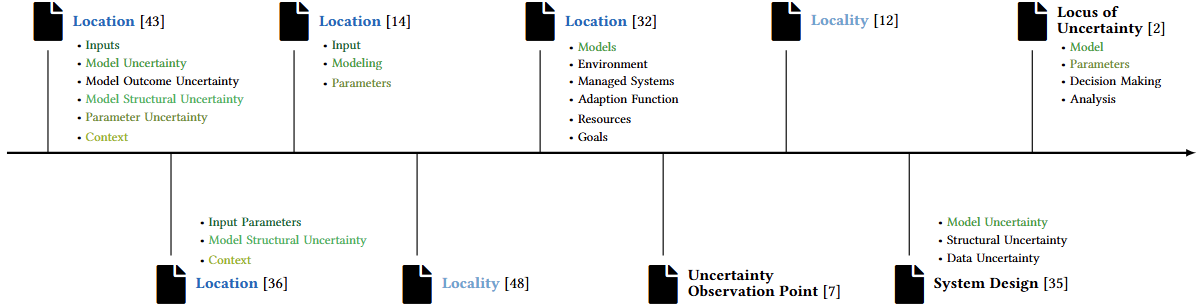}
    }
\caption{Overview of terms describing the Location of Uncertainty across various frameworks. Colors indicate terms with identical names but differing meanings across frameworks.}
\label{fig:uncertaintyLocation}
\end{figure*}

Our literature review identified several gaps and limitations within the existing research, which we explicitly describe in the following section:

\textit{\textbf{Gap 1:} - Jungle of Terminology Limiting Conceptual Clarity and Framework Integration:} \\
Our literature review revealed that a significant gap lies in the considerable variation in terminology across different frameworks.
%A significant gap is the considerable variation in the terminology across different frameworks. 
Many papers use different terms for the same concepts and, conversely, the same terms for different concepts.
These inconsistencies create a jungle of terminology, making it difficult to connect, align, and compare approaches effectively. 
%Figure~\ref{fig:uncertaintyLocation} presents a timeline illustrating the diverse terms used to describe the \textit{Location of Uncertainty}, highlighting the terminological inconsistencies identified in our literature review.
Figure~\ref{fig:uncertaintyLocation} presents a timeline illustrating how the terms used to describe the \textit{Location of Uncertainty} have evolved over time, highlighting the terminological inconsistencies identified in our literature review.
%Figure~\ref{fig:uncertaintyLocation} illustrates the diverse terms used to describe the \textit{Location of Uncertainty}, highlighting the terminological inconsistencies identified in our literature review.
%In total, nine different definitions and \mac{five different names, marked in gray boxes,} of the Location of Uncertainty are used across the related literature.
In total, nine different definitions and five different names of the Location of Uncertainty are used across the related literature.
%In Figure~\ref{fig:uncertaintyLocation} all reused category names are highlighted in green.
In Figure~\ref{fig:uncertaintyLocation}, all the reused category names are highlighted in the same blue tone.
%Across these nine definitions, 26 locations are identified, with 15 locations having different names, also marked in gray boxes, and 24 differing in meaning, marked with a star in the top right corner.
Across these nine definitions, a total of 26 locations are identified, with 15 having different names and 24 varying in meaning.
%Similar to the category names, all reused locations are marked in green, while those sharing the same meaning as another location are distinguished with a star in the top right corner.
Similar to the category names, all the locations with the same name but different meanings are marked in the same green tone.
%While those with the same meaning but different names are marked with the same symbolic representation. 
%Table~\ref{tab:terminologyStatistics} contains the full Statistics of the different terms and different meanings. 
%\input{tables/terminologyStatistics}
This variation highlights the significant complexity associated with just one category.
Additionally, such variability can lead to misunderstandings and complicate efforts to establish a unified framework, as readers and researchers must navigate this \textit{jungle of terminology} before making comparisons.

\textit{\textbf{Gap 2:} - Lack of Explicit Component Differentiation in CPS Frameworks:}\\%Component Differentiation in CPS Frameworks: 
An additional gap identified in our literature review is the absence of a CPS framework that explicitly differentiates between the distinct component types: the Cyber Component, the Physical Component, and the Platform Component. 
%\mac{and the Interaction between them}. 
While some works, such as those by Amsat et al.~\cite{asmat2023uncertainty}, Acosta et al.~\cite{acosta2022uncertainty}, and Zhang et al.~\cite{zhang2016understanding}, explore the domain of CPS, none provide a comprehensive framework that addresses these distinct components.

\textit{\textbf{Gap 3:} - Lack of Explicit Autonomy Differentiation in CPS Frameworks:} \\%Autonomy Differentiation in CPS Frameworks: 
The final gap we identified in our literature review is the absence of a framework that differentiates between autonomous and non-autonomous components within a system. 
This distinction is essential, particularly for autonomous CPS like self-driving cars, where some components function autonomously while others depend on external control or oversight. 
Without this distinction, representing, quantifying, and mitigating the Uncertainties in CPS becomes challenging, as each component type contains a different kind of uncertainty, and therefore, various techniques for quantifying and mitigating need to be applied. 
Without this differentiation, it becomes challenging to represent and quantify uncertainties, as each component type introduces distinct forms of uncertainty, requiring tailored techniques for both quantification and mitigation.

% Parts to rescue from PSUM 
To fill these gaps, we need a harmonizing framework that integrates the different definitions and proposals from the literature into a single representation. 
We have identified the PSUM Metamodel~\cite{PSUM2023} as a starting point to build such framework, given  %framework to harmonize diverse approaches and address the terminology jungle. %, forming the foundation of our strategy. 
%Our focus within PSUM is on 
its detailed uncertainty characterizations -- Effect, Uncertainty Perspective, and Pattern -- and its uncertainty attributes: Uncertainty Kind, Level, and Nature. 
Together, these six characteristics provide a good foundation for a more detailed description of uncertainties, which we will further extend to incorporate concepts relevant in the domain of CPS.
%- jungle of terminlogy 
%- different terms for the same thing 
%- more concrete highlights about things to reuse from PSUM 

\section{Harmonizing and Extended Conceptual Model}
\label{sec:proposedFramework}
%\jm{Include the meaning of the colors in Figure 3 }
Building on the gaps and limitations identified in Section~\ref{sec:gapsLimitations}, we present our Model for Representing Uncertainty to address these challenges. 
%This section introduces our Model for Uncertainty Representation, addressing the gaps and limitations identified in Section~\ref{sec:gapsLimitations}. 
Figure~\ref{fig:uncertaintyRepresentation} provides an overview of the model. 
Concepts shown in white are adopted from the PSUM Metamodel~\cite{PSUM2023}, those in blue originate from existing literature discussed in Section~\ref{sec:literatureReview}, and the green-colored concepts represent novel contributions introduced in this work.
%\textcolor{red}{(We may need to talk about the aggregation in Component called \texttt{AtomicComponents}, the UncertaintySourceType called \emph{Cyber-Physical Systems}) and the triangular relation between Uncertainty, Component and UncertaintyLocation}
%\begin{landscape}
%\begin{figure*}[t!]
%\centering
%    \includegraphics[width=0.85\linewidth]{figures/UncertaintyRepresentationBig.pdf}
%\caption{Harmonized and Extended Conceptual Uncertainty Representation Model.}
%\label{fig:uncertaintyRepresentation}
%\end{figure*}
%\end{landscape}

We address the limitations of existing uncertainty categorizations by providing a unified definition of each category, along with its available options and connections to other categories in Sections~\ref{sec:ConceptKind} - \ref{sec:ConceptSource}.
%First, we describe the categories adopted from the literature in Sections~\ref{sec:ConceptKind} - \ref{sec:ConceptSource}. 
%Each category is defined, accompanied by its available options and connections to other categories. 
Additionally, we introduce novel categories to address the limitations identified in the current uncertainty categorizations.
%Additionally, we address the identified gaps in the uncertainty categorization by introducing novel categories. 
In Section~\ref{sec:ConceptCPS}, we present categories that partition the complete CPS into smaller components, enabling a fine-grained characterization of uncertainty within the CPS.
%\mac{more precise localization} within the CPS. 
In Section~\ref{sec:ConceptAutonomy}, we introduce a characteristic that differentiates between autonomous and non-autonomous components.
%Finally, we highlight the previously missing distinction between Cyber-Physical Components in Sections~\ref{sec:ConceptComponent}-~\ref{sec:ConceptSystem} and Non-/Autonomous Components in Section~\ref{sec:ConceptAutonomy}.
\subsection{Uncertainty Kind}\label{sec:ConceptKind}
To classify the manifold types of uncertainties in the real world, we take the name of the category \textit{Uncertainty Kind} and its position as one of the three uncertainty attributes from the PSUM Metamodel~\cite{PSUM2023}.
%To describe the different types of uncertainty, we use the framework of the PSUM Metamodel~\cite{PSUM2023}. % terminology of the PSUM Metamodel~\cite{PSUM2023}. 
%We refer to this category as \textit{Uncertainty Kind}, treating it as one of the attributes of uncertainty.
For specific types, we incorporate the most relevant kinds identified by Troya et al.~\cite{troya2021uncertainty}, including Belief-, Occurrence-, Behavior-, Measurement-, and Spatiotemporal-Uncertainty.
The Spatiotemporal Uncertainty category combines Geographical and Time Uncertainty from Zhang et al.~\cite{zhang2016understanding}, to represent uncertainty about either the geographical or physical location or the time-related properties expressed in statements.
Occurrence Uncertainty extends the concept presented by Zhang et al.~\cite{zhang2016understanding}, broadening its focus from the degree of belief in the existence of an event to include the existence of an entity.
Behavior Uncertainty concerns the lack of knowledge about the behavior of the system or environment, including its actions, motivations, parameters, and timing.
Measurement Uncertainty refers to the range of possible states or outcomes of a measurement, with probabilities assigned to each.
Lastly, Belief Uncertainty addresses situations where the Belief Agent is uncertain about a statement, representing a form of second-order uncertainty.
\subsection{Reducibility}\label{sec:ConceptReducibility}
Understanding whether uncertainty can be reduced plays a critical role in determining how it should be addressed. 
To capture this dimension, we adopt the Reducibility Level category from the PSUM Metamodel~\cite{PSUM2023}. 
Within this framework, Reducibility is defined as an attribute of uncertainty and is classified into three levels: Fully Reducible, Partially Reducible, and Irreducible.

Fully Reducible Uncertainty refers to scenarios where uncertainty can be entirely eliminated, resulting in complete knowledge.
Partially Reducible Uncertainty applies to cases where uncertainty can be reduced, but only to a limited extent.
Irreducible Uncertainty describes situations where no reduction is possible, leaving uncertainty unavoidable.

We also introduce an additional category, Unknown, to capture cases where the reducibility of uncertainty cannot be determined.

%This classification allows CPS developers to prioritize efforts on addressing reducible uncertainties while preparing for irreducible ones, facilitating more effective decision-making and resource allocation.

\subsection{Nature}\label{sec:ConceptNature}
Following the PSUM Metamodel framework~\cite{PSUM2023}, the final uncertainty attribute is the Nature of Uncertainty.
The literature consistently defines this category and distinguishes between two types: epistemic and aleatory.
Epistemic uncertainty arises from imperfections in acquired knowledge, while aleatory uncertainty results from the inherent variability of the phenomenon being described.

\subsection{Effect}\label{sec:ConceptEffect}
%\jm{Include Fuzzy, interval, etc as possible extensions}
Once uncertainty is present in a system, %it influences the behavior of the system.
it can have a significant impact on the system's behavior, influencing its operations, outcomes, and decision-making processes.
This influence can manifest in various ways, all of which are captured in the category Uncertainty Effect.

In alignment with the PSUM Metamodel framework~\cite{PSUM2023}, we treat the Uncertainty Effect as a characteristic of uncertainty.
To represent this category, we adopt the approach proposed by Chipman et al.~\cite{chipman2015coverage}. 
Their framework introduces four distinct categories, each offering mathematical methods to quantify the impact of uncertainty.
These categories incorporate all relevant concepts identified in the literature and are structured around two key distinctions: continuous vs. discrete and non-deterministic vs. probabilistic.
The continuous probabilistic category includes representations such as random variables and probability distributions. %, which are among the most established and used methods to represent uncertainty. 
By contrast, the continuous non-deterministic category captures uncertainties using bounded but unspecified values.
This category can be further extended through approaches such as fuzzy set theory, possibility theory, and interval analysis. %, which offer refined means of representing vagueness, imprecision, or incomplete knowledge. 
The discrete non-deterministic category captures uncertainty through non-deterministic transitions between states, as represented in non-deterministic automata or transition systems. 
In contrast, the discrete probabilistic category models uncertainty by assigning probabilities to state transitions, as exemplified by probabilistic automata and Markov processes.

%The continuous probabilistic category includes random variables, while the discrete probabilistic category models probabilities in state transitions, such as probabilistic automata and Markov processes.
%In contrast, non-deterministic discrete uncertainties are represented using non-determinism in automata, whereas continuous non-deterministic uncertainties are modeled through ranges and intervals.
\subsection{Perspective}\label{sec:ConceptPerspective}
Uncertainty can be viewed from different perspectives.
For example, some uncertainties are identifiable through data analysis, while others can only be recognized through personal observations.
To capture this distinction, we adopt the Uncertainty Perspective category from the PSUM Metamodel~\cite{PSUM2023} as the next characteristic of uncertainty.
%The next Uncertainty characteristic we adopt from the PSUM Metamodel~\cite{PSUM2023} is the category of uncertainty perspective. 
%This category reflects the viewpoint of the observing agent.
A subjective perspective refers to uncertainty arising from the observations and reasoning of an agent, whereas an objective perspective remains independent of any particular observer.
\begin{landscape}
\begin{figure*}[t!]
\centering
    \includegraphics[width=0.85\linewidth]{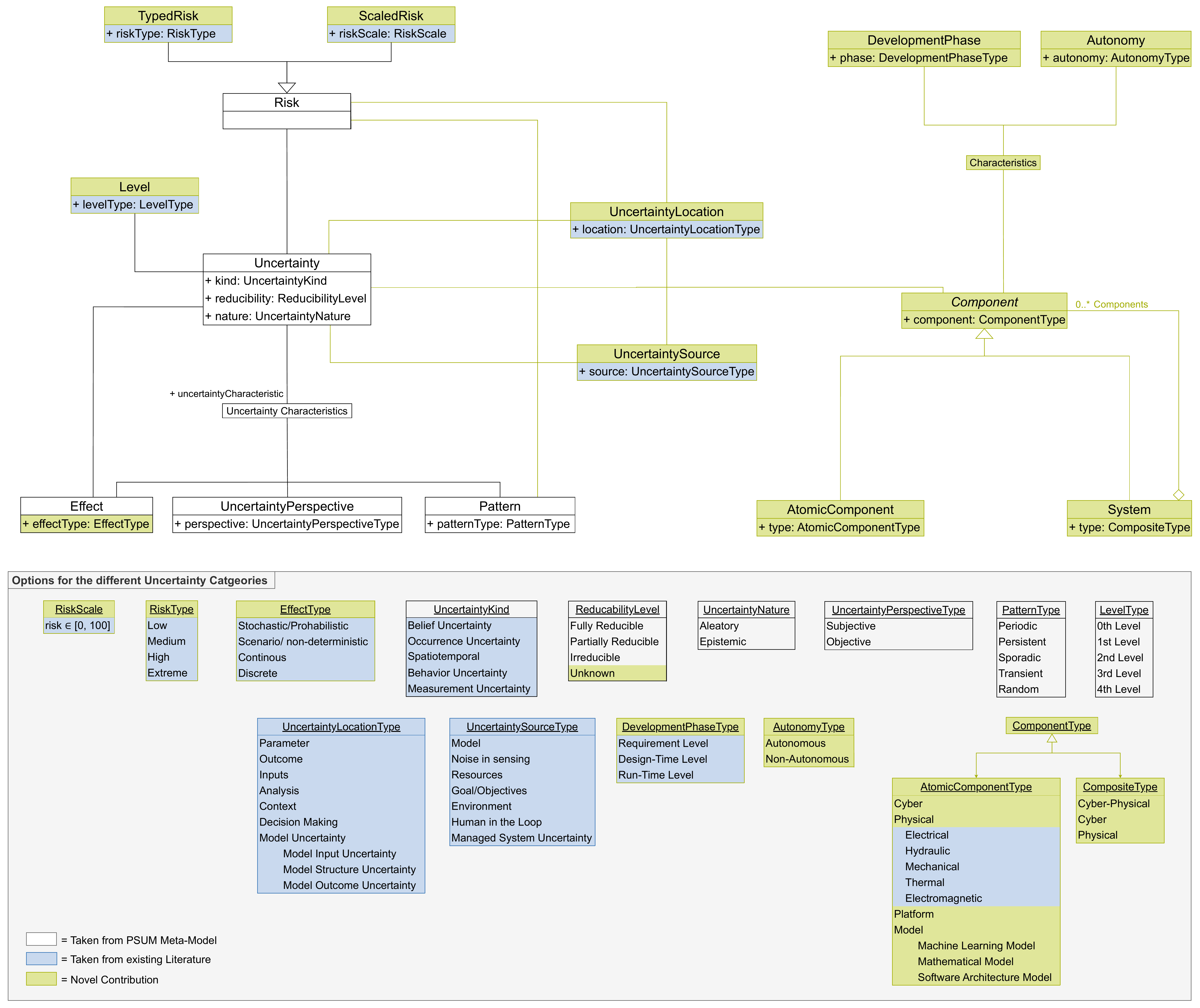}
\caption{Harmonized and Extended Conceptual Uncertainty Representation Model.}
\label{fig:uncertaintyRepresentation}
\end{figure*}
\end{landscape}
\subsection{Pattern}\label{sec:ConceptPattern}
Once it is established that uncertainty is present in a CPS, an important question is when this uncertainty arises.
To address this, we use the Uncertainty Pattern category, which is the final uncertainty characteristic in the PSUM Metamodel~\cite{PSUM2023}.
The Pattern of Uncertainty provides a framework for understanding how uncertainty manifests over time. 
To capture the various patterns of uncertainty occurrence, we adopt the classifications proposed by Zhang et al.~\cite{zhang2016understanding}.
Among these, only the random type does not follow a pattern
%\mac{only the occurrence pattern does not follow a pattern}
; the others follow temporal patterns.
The Periodic pattern occurs at regular intervals of time, while the Persistent pattern lasts indefinitely. 
Sporadic uncertainty appears occasionally (e.g., once under rare conditions during operation),
%\jm{ (a car’s keyless entry sometimes fails due to rare signal interference)}
and Transient uncertainty occurs temporarily (e.g., a short period under specific conditions before disappearing).
%\jm{ (a car’s touchscreen becomes unresponsive in extreme heat but functions normally after cooling)}. 
\subsection{Level}\label{sec:ConceptLevel}
Not all uncertainties have the same level of knowledge associated with them.
Some are well understood, while others lack sufficient knowledge or methods for discovery. 
To address these differences and better manage uncertainty, we classify them into levels.
We adopted the approach proposed by Perez Palacin~\cite{perez2014uncertainties}, as it captures the core ideas of other approaches in a clear and accessible manner.
Instead of using the term "orders of uncertainty" from the original framework, we refer to these categories as levels of uncertainty.
%The level of uncertainty reflects how much knowledge is available about the uncertainty, and, consequently, the extent to which it can be managed. 
%More knowledge allows for more effective management and mitigation, while limited knowledge can restrict the ability to address the uncertainty. To represent
%To represent these different levels,  we adopted the approach proposed by Perez Palacin~\cite{perez2014uncertainties}, as it captures the core ideas of other approaches in a clear and accessible manner.
%While Perez Palacin~\cite{perez2014uncertainties} uses the term "orders of uncertainty," we chose to refer to these as levels of uncertainty. 
The framework distinguishes five levels of uncertainty.
Level 0 represents the absence of uncertainty, signifying complete knowledge and serving as the ideal state.
Level 1 describes a recognized lack of knowledge, commonly referred to as known uncertainty.
Level 2 accounts for situations where there is both a lack of knowledge and a lack of awareness, which is identified as unknown uncertainty. 
Level 3 represents a state where no process exists to uncover the lack of awareness, making it impossible, with current possibilities, to transition from unawareness to recognizing uncertainty.
%In level 3, there is a lack of process to find out the lack of awareness. 
This means that with the current possibilities, it is not possible to move from not knowing to being aware of uncertainty. 
Finally, Level 4 addresses uncertainty about the levels themselves, referred to as Meta Uncertainty
\subsection{Risk}\label{sec:ConceptRisk}
When uncertainty is introduced into a system, it can significantly affect its behavior, influencing its operations, outcomes, and decision-making processes.
Given that systems often rely on expected functionality, any deviation in one system’s functionality can impact others and, in the worst case, lead to system failure.
Thus, uncertainty inherently introduces risk.
Since the risk varies depending on the uncertainty, we incorporate two complementary metrics to assess the risk induced by uncertainty: Risk Type and Risk Scale.
%In systems that involve uncertainty, it is essential to evaluate how severe this uncertainty is to manage its impact effectively. 
%To achieve this, we incorporate two complementary metrics to address the risk produced by the uncertainty: Risk Type and Risk Scale. 
The Risk Type is adapted from Zhang et al.~\cite{zhang2016understanding}, which builds upon ISO 31000~\cite{iso31000} to classify uncertainties into four levels: low, medium, high, and extreme.
This classification reflects the likelihood of the uncertainty occurring and its potential impact.
In addition, we introduce a novel numerical Risk Scale, ranging from 0 to 100. 
This scale allows for quantifying uncertainties and supporting their propagation and mitigation in future analyses.
\subsection{Location}\label{sec:ConceptLocation}
Understanding the location within a system where uncertainty is observed is important, as it provides valuable context for analyzing and managing the uncertainty effectively. 
For instance, the location of uncertainty can significantly influence the associated risk.
To address this, %To describe the location where uncertainty is observed, 
we identified the framework of Walker et al.~\cite{walker2003defining} and Acosta et al.~\cite{acosta2022uncertainty} as the most suitable for representing uncertainty locations in CPS.
Their focus on categorization through decision-based and coupled models highlights key aspects that significantly influence the CPS development.
Based on these frameworks, we define seven  categories: \textit{Parameter}, \textit{Outcome}, \textit{Inputs}, \textit{Analysis}, \textit{Context}, \textit{Decision Making}, and \textit{Model Uncertainty} which can be further split up into \textit{Model Input Uncertainty} and \textit{Model Structure Uncertainty}.

For Parameter Uncertainty, we adopt the definition provided by Walker et al.~\cite{walker2003defining}, which describes it as uncertainty arising from the data and methods used to calibrate model parameters.
Outcome Uncertainty is the uncertainty associated with the difference between a model’s outcome and the actual value, particularly when this difference is significant to the decision-maker, following Walker et al.~\cite{walker2003defining}.
For Input Uncertainty, we again refer to Walker et al.~\cite{walker2003defining}, who define it as the uncertainty related to the description of the reference system -- typically the current system -- and the external forces driving its changes.
The next category, Analysis Uncertainty, concerns uncertainties that arise from evaluating design decisions using model-based methods. 
This aligns with the definition provided by Acosta et al.~\cite{acosta2022uncertainty}.
Context Uncertainty addresses uncertainties related to the system's boundaries, focusing on what is included or excluded and the comprehensiveness of the system's representation of the real world. 
This understanding is consistent with Walker et al.~\cite{walker2003defining}.
From Acosta et al.~\cite{acosta2022uncertainty}, we adopt the category Decision-Making Uncertainty, which relates to uncertainties in the decision-making process, particularly when decisions are based on incomplete information or estimates of values of interest.
Finally, Model Uncertainty relates to uncertainties inherent to the model itself. In line with Walker et al.~\cite{walker2003defining}, this category is divided into two subcategories: Structural Uncertainty, concerning the form or structure of the model, and Technical Uncertainty, which refers to uncertainties in the model's computer implementation.
\subsection{Source}
\label{sec:ConceptSource}
Uncertainty Location refers to the part of a system where uncertainty is observed. 
Yet, the observed location is not always the root cause of the uncertainty. 
To address this distinction, we introduce the category of Uncertainty Source, which represents the origin of the uncertainty. 
The relationship between sources and locations is not one-to-one: a single Uncertainty Source can affect multiple locations, and conversely, a single Uncertainty Location can be influenced by multiple sources.

%Since the Location of Uncertainty refers to where uncertainty is observed within a system rather than its root cause, we introduce the category Uncertainty Source to address this distinction. 
%In this context, uncertainty arises from an Uncertainty Source and manifests at a specific location, referred to as the Uncertainty Location.\\
We highlight a selection of common sources of uncertainty, or classes of such sources, primarily based on the classification by Mahdavi-Hezavehi et al.~\cite{mahdavi2017classification}. 
This list is not exhaustive, as identifying all sources is domain-specific and remains future work. 
Tailored to CPS in general, we focus on the following five sources: Model, Resources, Goal/Objectives, Environment, and Managed System Uncertainty, all based on the descriptions by  Mahdavi-Hezavehi et al.~\cite{mahdavi2017classification}.
The class Model addresses uncertainties arising from model simplifications, incompleteness, drift, conflicting information, and complexity.
%
%Noise in sensing  
The Resources class relates to the system's resources, including uncertainties arising from the addition of new resources or the dynamic availability of services.
Uncertainties in the Goals/Objectives class relates to the goals or objectives of the system and includes ambiguities in goal definitions, conflicting objectives, and the potential for goals to evolve over time.
Environment uncertainties arise from external factors such as the execution context or human involvement.
In contrast,  uncertainties related to the system itself, such as complexity and dynamic changes, are categorized under the Managed System class.
%Cyber Physical Systems 

{\subsection{Relating uncertainty to CPS components}\label{sec:ConceptCPS}
%\diego{\subsection{Relating uncertainty to CPS components}}\label{sec:ConceptCPS}
%\textcolor{red}{Please check the following text. We can replace/merge the next section with this}

Identifying the part of the system where uncertainty manifests and is observed, i.e., the Uncertainty Location, as well as the source from which it arises, i.e., the Uncertainty Source, requires a concrete description of the relevant parts in the CPS rather than viewing the CPS as a whole.

To effectively represent the uncertainty aspects in concrete CPS parts while engineers can keep their autonomy to decide the appropriate modeling granularity for each CPS part, we apply the Composite pattern~\cite{patterns94}. The types of elements that represent the CPS following the composite pattern are called the Atomic Component, which is the finest modeling granularity in the CPS, and the System, which is the composite entity that can contain other Systems and Atomic Components. 

\subsubsection{Atomic Component}\label{sec:ConceptAtomicComponent}
An atomic component represents the finest granularity unit of a CPS considered in our representation model.
For CPS, the possible atomic component types are: Cyber, Physical, Platform parts, and the Model type. %\textcolor{red}{Say something about the Model Type because it appears in the conceptual model in the figure.}

The Physical Part of a CPS includes components that interact directly with the environment, enabling the system to engage with physical processes. 
To refine the Physical Part further, we follow the approach proposed by Karsai~\cite{karsai2015modeling} and classify it into the following categories: Electrical, Hydraulic, Mechanical, Thermal, and Electromagnetic.

The Cyber Part refers to the digital and computational components responsible for processing, analysis, decision-making, and control within the system.

A critical aspect of a CPS is the Platform Part, which facilitates interaction between the two other components. 
This interaction often introduces additional complexity, and thus potential uncertainty, into the system. 
To emphasize its significance, the interaction platform is treated as a separate atomic component type, as it plays a key role in influencing overall system behavior and uncertainty propagation.

Our model also allows the representation of the uncertainty related to the various types of models used in the CPS.
The types of models include Mathematical Models, Software Architecture Models, and Machine Learning Models, which represent diverse approaches to analyzing and understanding system behavior. 
It is important to note that this list reflects current observations from the reviewed literature but remains incomplete and open to future extensions.

\subsubsection{System}\label{sec:ConceptSystem}
While atomic components provide the fundamental building blocks of a CPS, some functionalities require more complex groupings of these building blocks. Hence, we introduce the System entity as the aggregator of other entities. 
This hierarchical structure of the Composite pattern allows a System to be viewed as a CPS in its own right. 
Consequently, we introduce a new composite type, the Cyber-Physical, to capture the combined nature of cyber and physical elements at the entity container level.
At the root level of the composite structure, the System entity represents the whole CPS, which is composed of other lower granularity Systems and or Atomic Components

\subsection{Development Phase}\label{sec:ConceptDevelopment} 
Following the literature, we also introduce a category that represents the time when uncertainties can arise. 
%Following the literature we also make a distinction between different times at which uncertainties can arise. 
To tailor this concept more specifically to CPS, we renamed this category the Development Phase of a CPS, as it captures the stages of development where uncertainties tend to occur more accurately. 
Following the differentiation proposed by Ramirez et al.~\cite{ramirez2012taxonomy}, we incorporate three development phases in our unified model: Requirement-Level, Design-Time, and Run-Time.
The Requirement-Level refers to the phase in which development requirements are established, Design-Time describes the phase during which development occurs, and Run-Time is the phase after the system has been deployed.
\subsection{Autonomy}
\label{sec:ConceptAutonomy}
CPS are increasingly integrating both autonomous and non-autonomous components. 
Since these parts have fundamentally different characteristics, uncertainties must be handled according to their respective status. 
To address this, differentiating between autonomous and non-autonomous parts becomes a crucial addition to uncertainty representation.
Autonomous parts are capable of making decisions independently, often relying on internal models, real-time sensor data, and adaptive mechanisms.
In contrast, non-autonomous parts operate based on explicit commands or inputs from external controllers.
We assign this characteristic to each component to enable flexibility in defining the autonomy level of system parts. 
This approach allows both atomic components and entire systems to be classified as either autonomous or non-autonomous.

\subsection{Summary of Contributions}
\label{sec:contribution}
%\mac{Maribel: Maybe we can include here one last subsection summarizing the novelties of our model, and how we fill the gaps identified in Section 2. [To discuss with everyone] }
This section summarizes how we addressed the gaps and limitations identified in  Section~\ref{sec:gapsLimitations} and highlights the novel parts of our harmonized conceptual model. 

\textit{\textbf{Contribution 1:} - Systematic Refinement of Terminology for Uncertainty Representation in CPS}
The first gap we identified is the inconsistency in terminology, where numerous overlapping or redundant terms complicate the understanding of uncertainty categorization. 
We tackled this issue by systematically analyzing each term and concept by identifying equivalences, overlaps, and hierarchical relationships.
This process ensured that only the essential terms and concepts remained while still preserving a comprehensive overview.
For instance, in the category of Uncertainty Location, Figure~\ref{fig:uncertaintyLocation} illustrates that existing literature provides nine different definitions and 26 different locations.
Through our analysis, we consolidated these into a single definition with 10 distinct locations, as illustrated in Figure~\ref{fig:uncertaintyLocationExtended}. 
The timeline in Figure~\ref{fig:uncertaintyLocation} has been extended to incorporate our proposed categorization, highlighted in orange, making the reduction in redundancy and the improved structure clearly visible.
\begin{figure*}[t!]
\centering
    \resizebox{0.95\textwidth}{!}{%
    \includegraphics{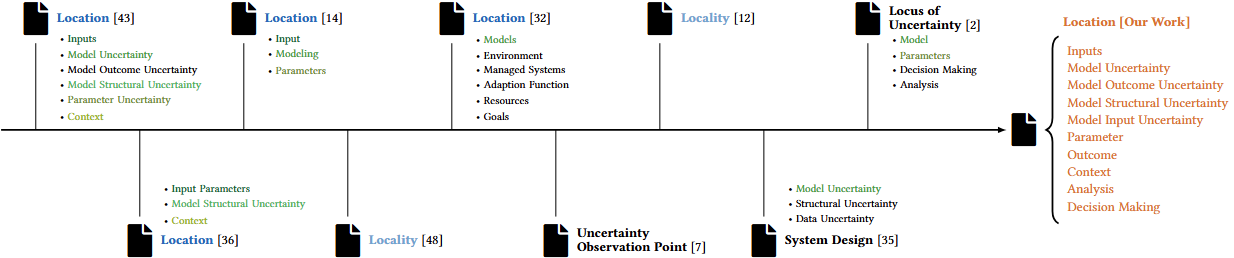}
    }
\caption{Overview of terms describing the Location of Uncertainty across various frameworks. Colors indicate terms with identical names but differing meanings across frameworks, while orange highlights the introduced consistent terminology. }
\label{fig:uncertaintyLocationExtended}
\end{figure*}

\textit{\textbf{Contribution 2:} - Structured Classification of CPS Components for Uncertainty Representation}\\
Addressing the gap of a systematic CPS component classification, Section~\ref{sec:ConceptCPS} introduces a structured framework that distinguishes between Cyber, Physical, and Platform components.
Unlike existing approaches that consider CPS as a monolithic system, our model introduces a hierarchical decomposition into Atomic Components, Components, and Systems. 
This structure enables precise localization of uncertainty by linking it directly to specific CPS elements rather than relying on a generic classification. 
Additionally, we introduce the Platform Component as a distinct entity to capture interactions between Cyber and Physical elements, a crucial yet often overlooked aspect in uncertainty representation.

\textit{\textbf{Contribution 3:} - Classification of Autonomous and Non-Autonomous Components in CPS} \\
The distinction between autonomous and non-autonomous components introduced in Section~\ref{sec:ConceptAutonomy} extends the CPS characterization and addresses the final gap identified in our literature review.
%Furthermore, we extend the characterization of CPS components by introducing in Section~\ref{sec:ConceptAutonomy} a distinction between autonomous and non-autonomous components.
Autonomous components operate independently using internal models and real-time data, while non-autonomous components rely on external commands. 
This refinement enhances uncertainty representation by capturing the differing decision-making properties within CPS.

\section{Conceptual Model Application Example}
\label{sec:example}

We illustrate the application of the conceptual model to a CPS, using the Autonomous Vehicle (AV) as a CPS example. Since the AV includes a multitude of components, we concentrate on the model of a set of components in the self-driving pipeline. Among other components, the self-driving system pipeline includes components for data \emph{Acquisition}, \emph{Perception} of objects, traffic signs, pedestrians, etc., and \emph{Actuation} on the vehicle physical devices such as steering, throttle, brakes, etc. We assume that the \emph{Perception} system is purely software, and therefore, the componentType is \emph{cyber}; the \emph{Acquisition} componentType is cyber-physical since it includes the physical devices that capture data from the real world as well as the digital representation of such data; while the \emph{Actuation} system components are purely physical. Figure \ref{fig:example-components} shows the representation of subsystems and atomic components of the AV instantiating the model in Figure \ref{fig:uncertaintyRepresentation}.

\begin{figure}[t!]
{\centering
    \includegraphics[width=0.8\textwidth]{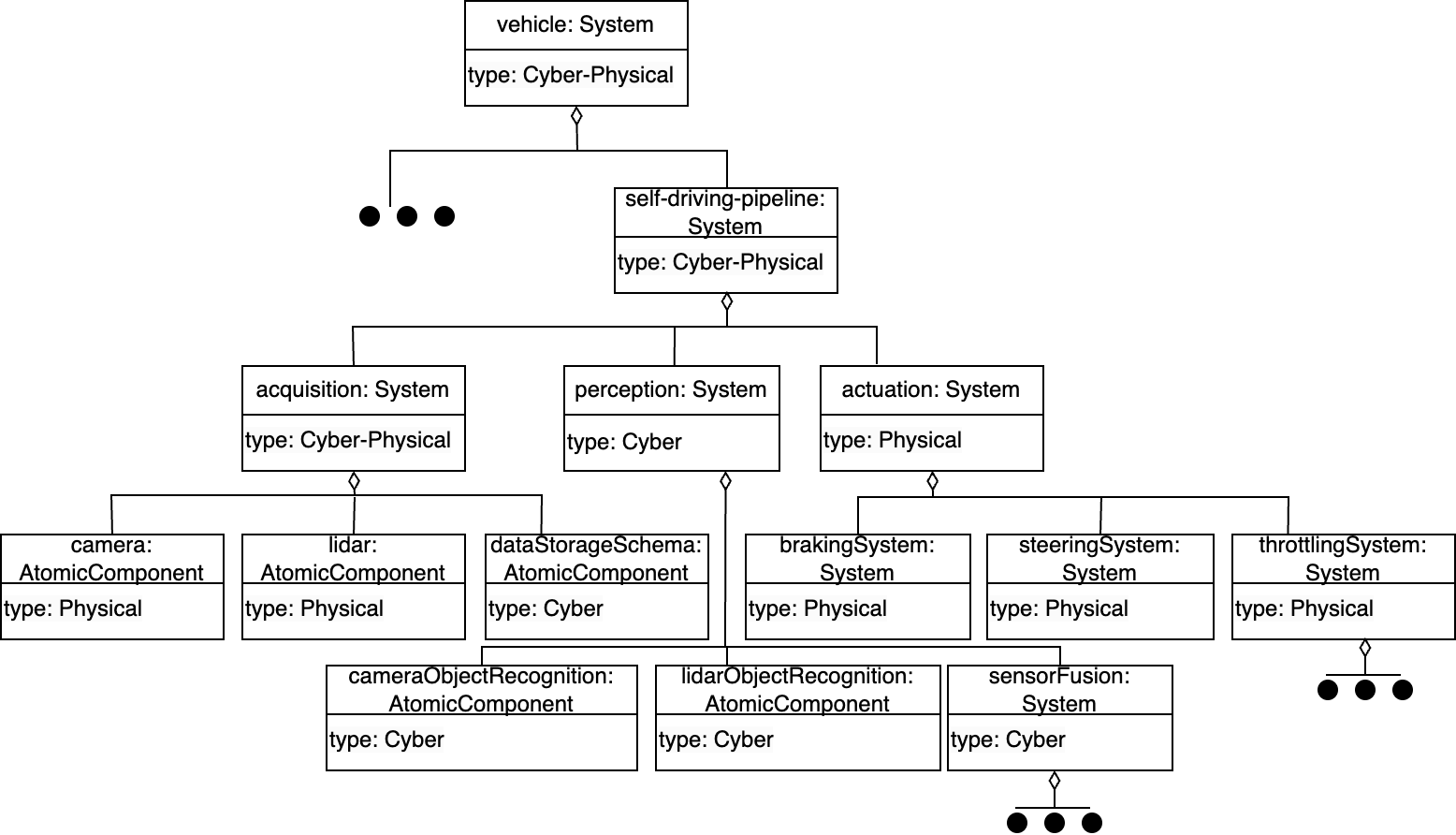}
    
\caption{Model of a selection of components in the autonomous vehicle}
\label{fig:example-components}
}
\end{figure}

Regarding the Autonomy associated with Components, The \emph{Actuation} and \emph{Perception} systems are \emph{Autonomous}. For vehicles that are not possible to actuate manually (e.g., autonomous delivery robots that are not possible to steer manually but move based on their defined destination), the \emph{Actuation} system is \emph{Autonomous} too. 
However, if we assume that the AV is one of the current cars with some self-driving capabilities, the \emph{Actuation} system is \emph{non-autonomous} since it demands oversight and external control sometimes. In that case, the \emph{non-autonomous} Autonomy of the \emph{Actuation} system propagates upstream in the composite structure \cite{patterns94}. 
Regarding the DevelopmentPhase, the illustrated example concentrates on the \emph{Run-Time level}.

Figures \ref{fig:example-camera} and \ref{fig:example-camera-recognition} illustrate the information of a few uncertainties associated with Components. Figure \ref{fig:example-camera} represents the uncertainty regarding the noise captured by the camera component. It models that it is a \emph{Measurement Uncertainty} since it comes from capturing information from reality, and hence its source of the uncertainty is the \emph{Noise in Sensing}. The location of the uncertainty lies in the camera component output. We assume that the camera is unaware of its uncertain behavior and does not provide any information about its noise in the images captured, then getting assigned a 2nd level. As typically happens with the noise, it is assumed to be aleatory, and this uncertainty about the reality values cannot be eliminated with more data. Since the components that use the camera as input are well aware of the existing noise when acquiring images with cameras, and they behave according to the existence of noise, the risk caused by the noise in sensing in the camera is considered \emph{Low}.

In turn, Figure \ref{fig:example-camera-recognition} shows two uncertainties associated with the cameraObjectRecognition Component. The first uncertainty, named \emph{u2} in the figure, refers to the imperfection of the algorithms that classify elements in images. Therefore, its kind is the \emph{Behavior Uncertainty}, and its location is the output of the object classification algorithms. Since it is possible to reduce the uncertainty in the classification by providing more training data or a better training set, the uncertainty reducibility is \emph{partially reducible} and its nature is \emph{epistemic}. We assign \emph{1st level} to this uncertainty because, in modern systems, there is awareness that automated classifications are not always perfect and mitigation techniques are implemented. For example, our example already considered a sensorFusion subsystem that combines results from the object recognition software and aims at mitigating the impact of their imperfect classifications. Since the sensorFusion exists, we assigned only a \emph{Medium} risk to the cameraObjectRecognition. However, if this program was the only authority to classify elements on the road, its assigned risk would have been higher. 
To show how a component may hold more than one uncertainty, the right part of the figure shows an additional associated uncertainty, called \emph{u3}. This uncertainty refers to the quality of the \emph{Inputs} that cameraObjectRecognition receives. Images may of of low quality due to the presence of dust on the camera, or because an element in the environment partially occludes the actual scene to process, or others. 

\begin{figure}[t!]
{\centering
    \includegraphics[width=0.65\textwidth]{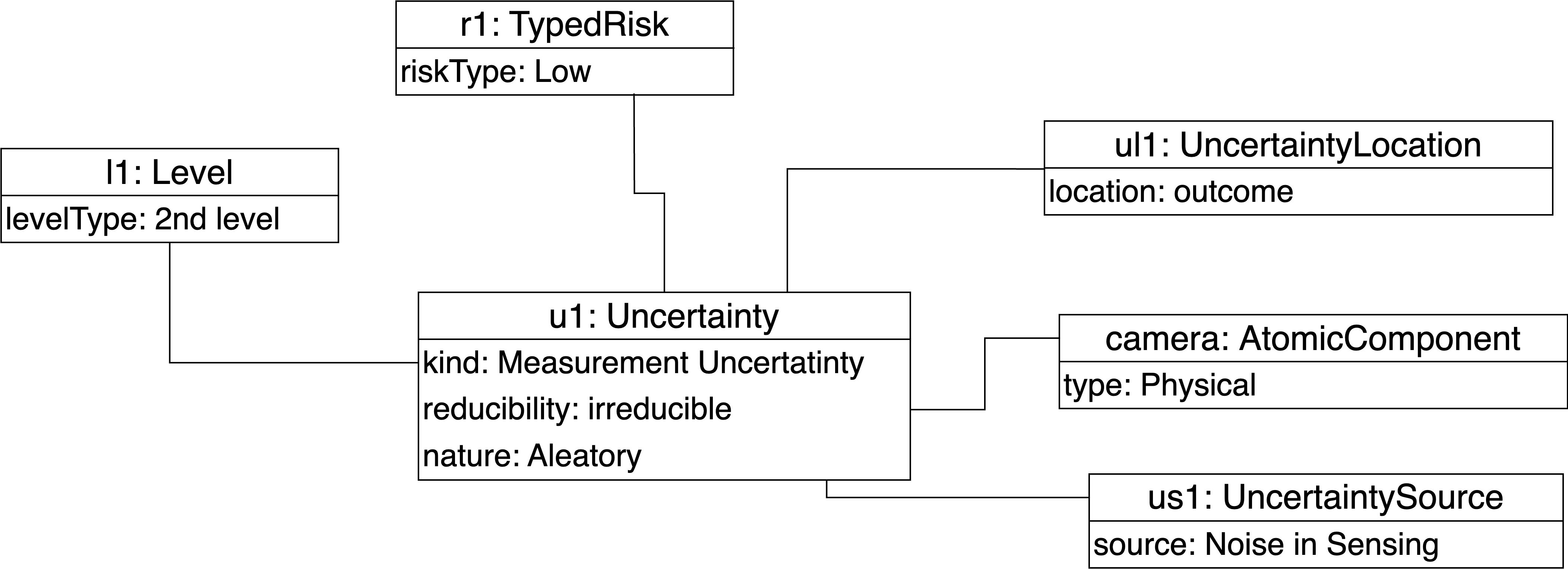}
\caption{Model of a selection of uncertainties in the camera component}
\label{fig:example-camera}
}
\end{figure}

\begin{figure}[t!]
{\centering
    \includegraphics[width=0.8\textwidth]{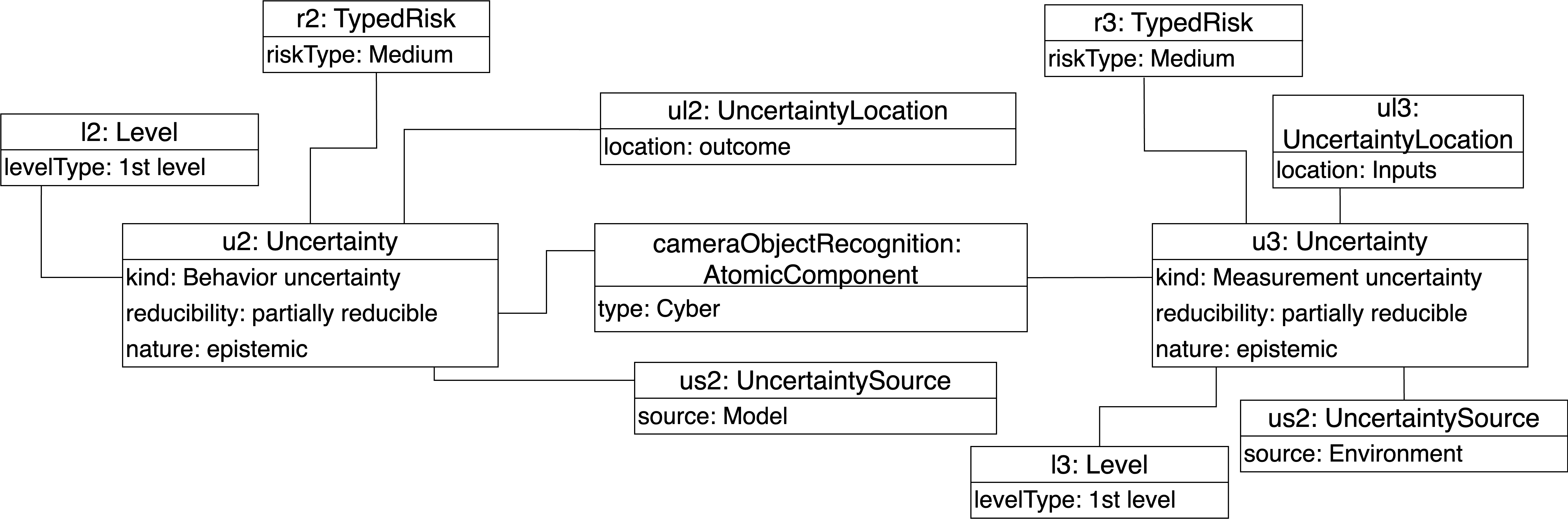}
\caption{Model of a selection of uncertainties in the object recognition from the camera information}
\label{fig:example-camera-recognition}
}
\end{figure}

\section{Conclusion and Future Directions}
\label{sec:conclusion}
In this paper, we have conducted a literature review of uncertainties in CPS and identified some gaps in the current classification schemes.
We then proposed a new uncertainty representation model that addresses limitations in previous classifications and tries to overcome the "jungle of terminology" problems. Our work has highlighted the multifaceted nature of uncertainties in CPS and paved the way for future research in this domain. 

\paragraph{Future Research Directions}
Based on our research, we conclude that incorporating the concept of uncertainty from the outset of software design is crucial for the engineering of future CPS. This approach would facilitate the development of CPS that are inherently aware of uncertainty, guided by a set of design principles and methodologies that intrinsically address various forms of unpredictability. Furthermore, this perspective opens up several promising research avenues that warrant exploration to achieve this overarching objective. These directions are briefly described below.

\textbf{Shared Understanding.} Our work is a first step towards a unified uncertainty terminology. Additional work in this direction is necessary to provide a common basis for uncertainty representation, measurement, and reporting across different CPS domains, facilitating better communication and comparison of results.

\textbf{Quantification methods.} The resulting model and the different facets of uncertainty highlighted emphasize the need to investigate and develop sophisticated techniques to quantify and measure various types of uncertainties in CPS, particularly those that are difficult to capture using traditional probabilistic approaches.

\textbf{Propagation and Interaction} The complexity of the CPS architecture requires investigation of how uncertainties propagate through the different components and how uncertainties in one domain (e.g., cyber) affect uncertainties in another (e.g., physical). This study would also lead to the development of new comprehensive mitigation strategies.

\textbf{Human-system interaction} The increasing diffusion of CPS and their usage together with humans requires an in-depth investigation of how human-system interactions affect and /or mitigate uncertainties in CPS. This, in turn, would require the development of methods for effective human-machine collaboration under uncertainty.

\section*{Acknowledgments}
Funded by the Deutsche Forschungsgemeinschaft (DFG, German Research Foundation) – SFB 1608 – 501798263

%Bibliography
\bibliographystyle{unsrt}  
\bibliography{references}

\end{document}